\documentclass[review]{elsarticle}
\makeatletter
\def\ps@pprintTitle{%
	\let\@oddhead\@empty
	\let\@evenhead\@empty
	\def\@oddfoot{\centerline{\thepage}}%
	\let\@evenfoot\@oddfoot}
\makeatother
\usepackage{lineno,hyperref}
\usepackage{graphicx}
\usepackage{listings}
\usepackage{amsmath}
\usepackage{amssymb}
\usepackage{tikz}
\usepackage{caption}
\usepackage{subcaption}
\usepackage{algorithm2e}
\usepackage{setspace}
\usepackage[title]{appendix}
\usetikzlibrary{automata, positioning, arrows, bayesnet, matrix,shapes,chains, calc}

\newtheorem{innercustomgeneric}{\customgenericname}
\providecommand{\customgenericname}{}
\newcommand{\newcustomtheorem}[2]{%
  \newenvironment{#1}[1]
  {%
   \renewcommand\customgenericname{#2}%
   \renewcommand\theinnercustomgeneric{##1}%
   \innercustomgeneric
  }
  {\endinnercustomgeneric}
}
\newcustomtheorem{alg}{Algorithm}

\newcommand{\vect}[1]{\ensuremath{\boldsymbol{\mathbf{#1}}}} % vector
\newcommand{\mat}[1]{\ensuremath{\boldsymbol{\mathbf{#1}}}} % matrix
\newcommand{\Cov}{\boldsymbol{\mathrm{Cov}}} % Covariance matrix
\newcommand{\MMAP}{{\mathrm{MMAP}}} 
\newcommand{\MAP}{{\mathrm{MAP}}} 
\newcommand{\argmax}{{\mathrm{argmax}}} % vector

\modulolinenumbers[5]

\journal{Spatial Statistics}

\begin{document}
\begin{frontmatter}

\title{Spatio-temporal Inversion using the  Selection Kalman Model}

 \author{Maxime Conjard\fnref{cor1}}
  \ead{maxime.conjard@ntnu.no}
  \author{Henning Omre}
  \ead{henning.omre@ntnu.no}
 	
\fntext[cor1]{Corresponding author}
\address{Department of Mathematical Sciences, NTNU, NO-7491 Trondheim, Norway}

\begin{abstract}
Data assimilation in models representing spatio-temporal phenomena poses a challenge, particularly if the spatial histogram of the variable appears with multiple modes. The traditional Kalman model is based on a Gaussian initial distribution and Gauss-linear dynamic and observation models. This model is contained in the class of Gaussian distribution and is therefore analytically tractable. It is however unsuitable for representing multimodality. We define the selection Kalman model that is based on a selection-Gaussian initial distribution and Gauss-linear dynamic and observation models. The selection-Gaussian distribution can be seen as a generalization  of the Gaussian distribution and may represent multimodality, skewness and peakedness. This selection Kalman model is contained in the class of selection-Gaussian distributions and therefore it is analytically tractable. An efficient recursive algorithm for assessing the selection Kalman model is specified. The synthetic case study  of spatio-temporal inversion of an initial state, inspired by pollution monitoring, containing an extreme event suggests that the use of the selection Kalman model  offers significant improvements compared to the traditional Kalman model when reconstructing  discontinuous initial states. 
\end{abstract}


\end{frontmatter}

%\linenumbers

\section{Introduction}
Data assimilation in models representing spatio-temporal phenomena is challenging. Examples can be found in pollution monitoring, weather forecast and petroleum engineering. In air pollution monitoring, see \cite{hopke}, potential source contribution (PSC) identification is an issue and inverse trajectory methods are often used to retrieve such maps. Source mapping from airborne smoke from wild fire  is one compelling example, see \cite{Cheng}. We present an alternative methodology which appears  as suitable for identification of extreme events as source for spatio-temporal phenomena. 
The identification of the source of the contamination in groundwater pollution can also be challenging.  Most studies focus on  the future distribution of the pollutant plume, but as \cite{kjeldsen} emphasize, the source will often be highly heterogeneous. Various Kalman type models are frequently used in hydrology, see \cite{ferra}, but identification of sources that appear as extreme events can be challenging. We believe that the Kalman type model defined in this study is suitable for source mapping  of these events. In petroleum reservoir characterization, multimodal  spatial histograms also appear due to spatially varying lithologies.  Assimilation of production data is then challenging, see \cite{Dovera}. We present an alternative model for these multimodal spatial variables. 

The traditional Kalman model as introduced by Kalman in his seminal paper \cite{kalman} provides a frequently used framework for evaluating spatio-temporal phenomena. It assumes Gauss-linear dynamic and observation models along with a Gaussian initial distribution. The Kalman model is therefore analytically tractable and is contained in the class of Gaussian distributions. As such the model is suitable to assess smooth spatial variables with linear dynamics and data collection. Various models stemming from Kalman's idea, for example the ensemble Kalman Filter \cite{evensen} and the unscented Kalman filter \cite{Julier}, are used to represent phenomena with non linear dynamics. Unfortunately, analytical tractability is lost for these models and the  distributions of interest are not contained in the class of Gaussian distributions. For spatial variables with spatial histograms that are  skewed or multimodal,  non-Gaussian initial distributions should be specified. Skew-Gaussian  spatial distributions are discussed in \cite{kim} and \cite{kjartan}, while a corresponding Kalman model is defined in \cite{naveau}. Multimodality in spatial variables is more complicated to represent. The ensemble Kalman Filter will fast regress towards a unimodal model due to the linearization of the observation conditioning. It is unclear how to adapt the unscented  Kalman filter to multimodal variables. A Gaussian mixture model in a spatial setting may be defined, but it must include mode indicators with spatial dependence, see \cite{kjartan2}. This latent categorical mode indicator complicates the definition of data assimilation in a Kalman framework. Alternatively, multimodal spatial variables can be represented by selection-Gaussian distributions, see \cite{delpino} and \cite{branco}, which appears as a generalization of the skew-Gaussian distribution. This selection-Gaussian spatial model may represent peaked, skewed and multimodal variables, see \cite{henning}. 
\\
We define a selection Kalman model with Gauss-linear dynamic and observation models and an initial model in the class of selection-Gaussian distributions. We demonstrate that the selection Kalman model is contained in the class of selection-Gaussian distributions and therefore it is analytically tractable. Since the Gaussian distribution appears as a central case in the class of selection-Gaussian distributions, the selection Kalman model may be seen as a generalization of the traditional Kalman model.
\\
In this paper $\vect{y} \sim f(\vect{y})$ denotes a random variable $\vect{y}$ distributed according to the probability density function (pdf) $f(\vect{y})$, or alternatively according to the corresponding cumulative distribution function (cdf) $F(\vect{y})$. Moreover, $\varphi_n(\vect{y};\vect{\mu},\mat{\Sigma})$ denotes the pdf of the Gaussian $n$-vector $\vect{y}$ with expectation $n$-vector $\vect{\mu}$ and covariance $(n \times n)$-matrix $\mat{\Sigma}$. Further $\Phi_n(A;\vect{\mu},\mat{\Sigma})$ denotes the probability of the aforementioned  Gaussian $n$-vector $\vect{y}$ to be in $A \subset \mathbb{R}^n$. We also use $\vect{i}_n$ to denote the all-ones $n$-vector and $\mat{I}_n$ to denote the identity $(n\times n)$-matrix. 
\\
In Section 2, the problem is set. In Section 3, the traditional Kalman model is cast in a Bayesian hidden Markov model framework. The generalization  to the selection Kalman model is then defined, and the analytical tractability is investigated. Further an efficient recursive algorithm for assessing the posterior distribution is specified. In Section 3, a synthetic case study  of the convection diffusion equation  is chosen to showcase the ability of the class of  selection-Gaussian distributions. The goal is to reconstruct the initial state which contains an extreme event. Results from the selection Kalman model and the traditional Kalman model are compared. In section 4, conclusions are presented.

\section{Problem Setting}
\begin{figure}
	\centering
	\includegraphics[width=0.5\textwidth]{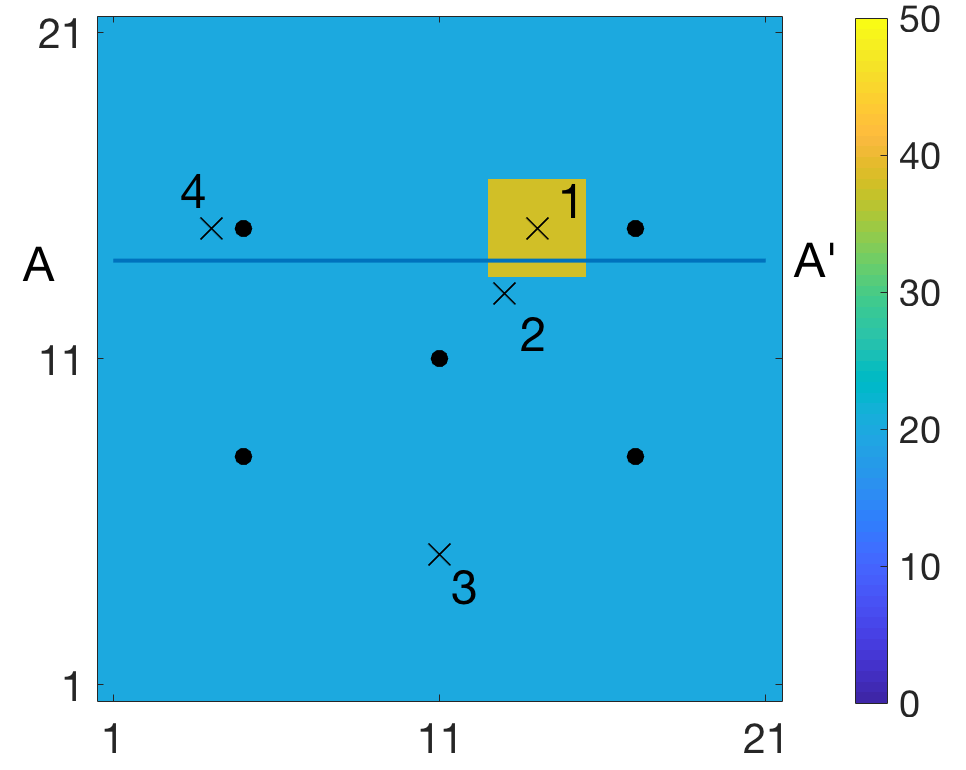}
	\caption{Initial state with observation locations ($\cdot$) and monitoring locations ($\times$)}
	\label{hm1}
\end{figure}
\label{section2}
\begin{figure}
	\centering
	\includegraphics[width=\textwidth]{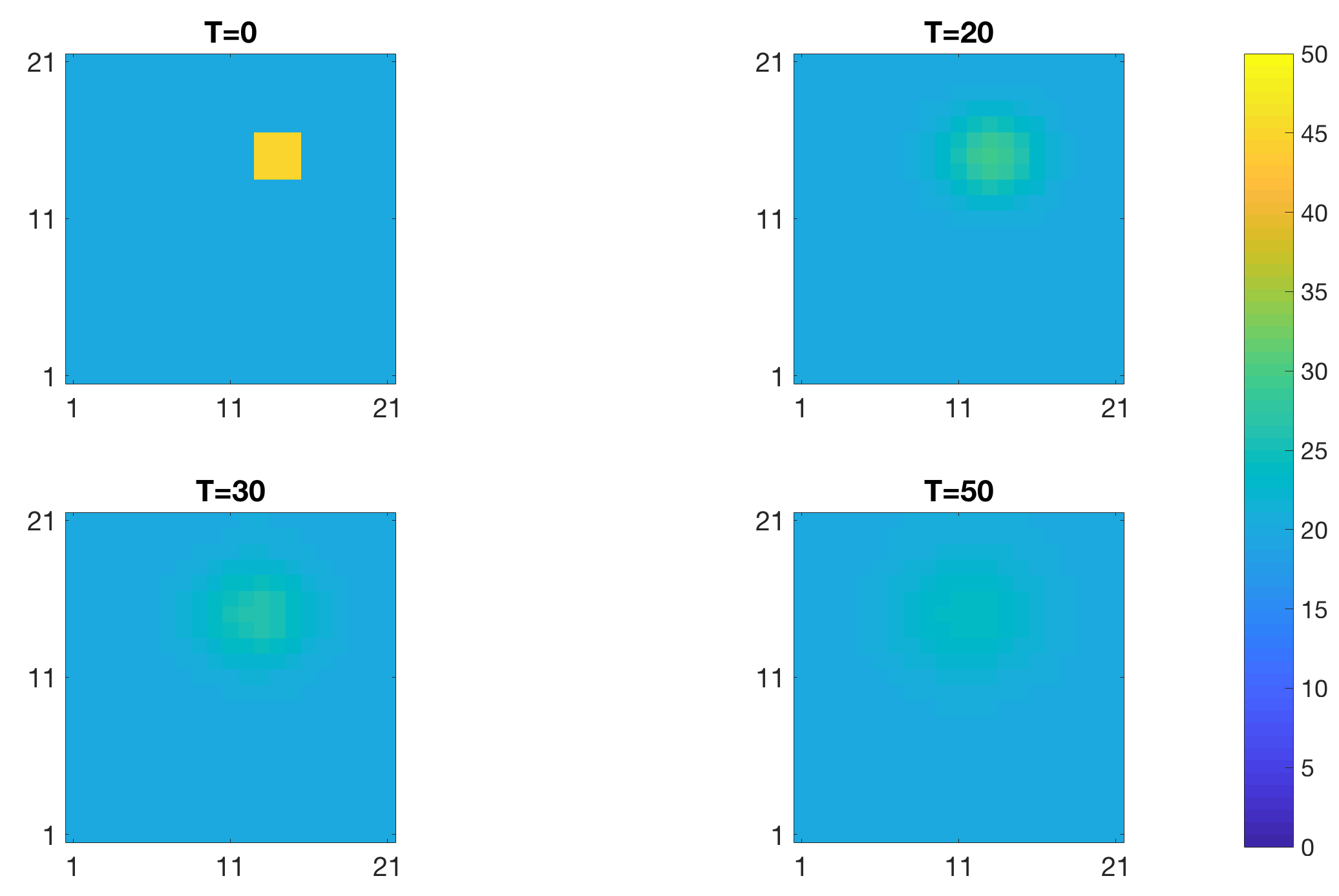}
	\caption{Spatio-temporal diffusion}
	\label{data}
\end{figure}
\begin{figure}
	\centering
	\includegraphics[width=0.6\textwidth]{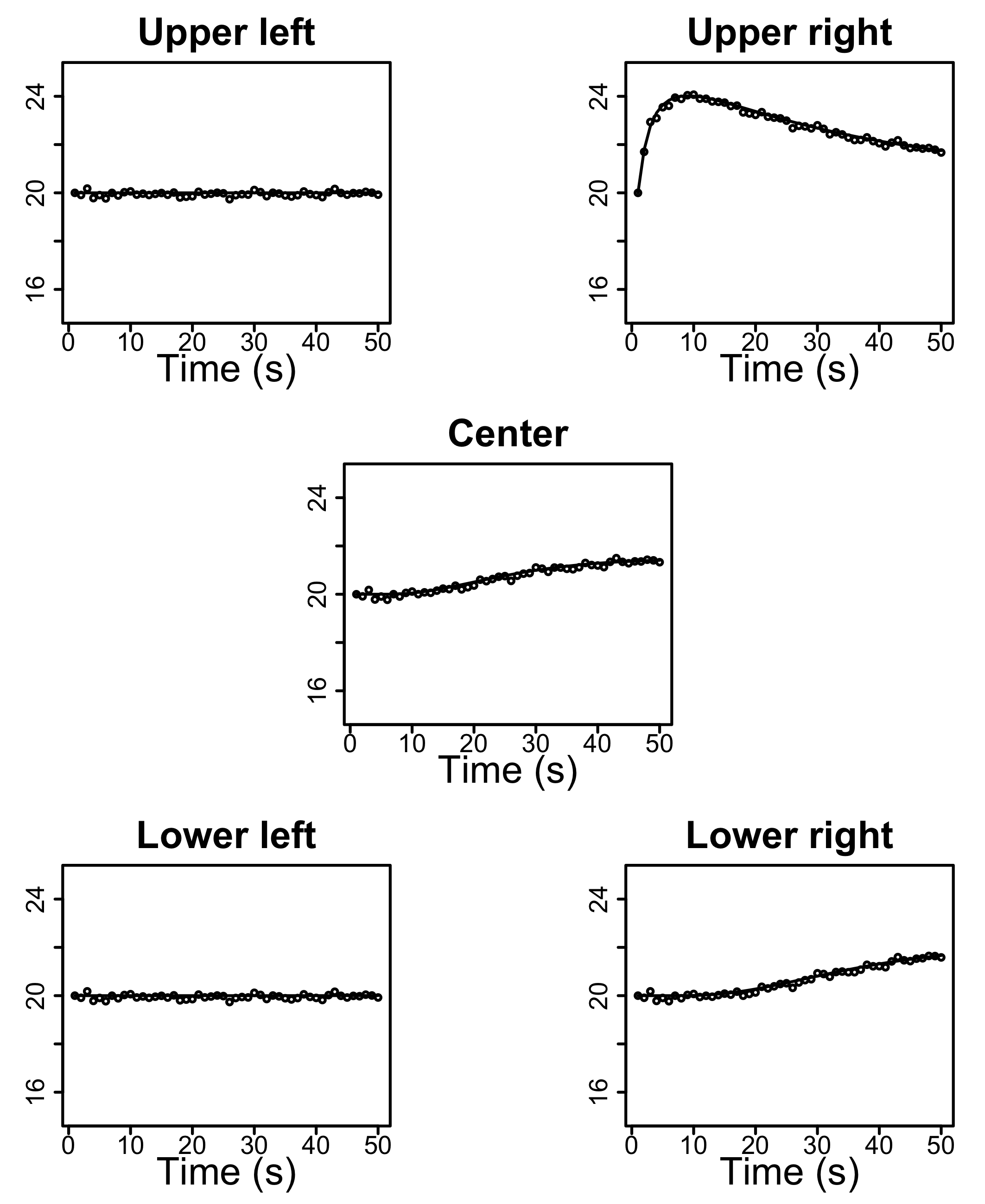}
	\caption{ Observations at the observation locations and true curve}
	\label{data2}
\end{figure}
The case is defined in a spatio-temporal setting. Consider the variable $\{ r_t(\vect{x}); \vect{x} \in \mathcal{L}_r, t \in \mathcal{T}\};r_{\cdot}(\cdot) \in \mathbb{R}$, with $\mathcal{L}_r$ a grid of size $n$ over a two-dimensional spatial area of interest while $\mathcal{T}:\{0,1,\ldots,T\}$ is a regular discretization in time. Let $t=T$ represent current time while $t=0$ represents the initial time. The spatial variable $\{ r_0(\vect{x}); \vect{x} \in \mathcal{L}_r\}$ is a discretized representation of the initial state which later will be assumed to be unknown. The initial state in the synthetic study is displayed in Figure \ref{hm1}. The state variable could for instance be temperature or the concentration of a pollutant, which does not vary significantly, bar an extreme event.\\
The spatio-temporal  variable evolves in time, $\{r_{t+1}(\vect{x}); \vect{x} \in \mathcal{L}_r\} = w_t[\{ r_t(\vect{x}); \vect{x} \in \mathcal{L}_r\}]$ where $\omega_t(\cdot)$ is a  dynamic function usually represented by a set of discretized differential equations and Figure \ref{data} displays the time development in this study. 
The spatio-temporal variable is not fully observable, it can only be measured at a number of monitoring sites. The actual observations have some measurement errors, and they appear as time series at the observation sites denoted $ \{\vect{d}_t = (\vect{d}_t^1,\ldots,\vect{d}_t^m), t \in \mathcal{T}\}$ where $m$ is the number of observation sites. The observations in the synthetic study are presented in Figure \ref{hm1} and \ref{data2}, the sites in the former and the actual time series in the latter. 
The  typical challenge  is to infer  the spatio-temporal variable $\{ r_t(\vect{x}); \vect{x} \in \mathcal{L}_r, t \in \mathcal{T}\}$ based on the observed time series $\{\vect{d}_t; t \in \mathcal{T}\}$. This challenge constitutes a complex spatio-temporal inverse problem. In the current study we focus on assessing the initial spatial variable $\{ r_0(\vect{x}); \vect{x} \in \mathcal{L}_r\}$  from the observed time series $\{ \vect{d}_t; t \in \mathcal{T}\}$.
\section{Model Definition}
Consider the unknown temporal $n$-vector $\vect{r}_t$, representing the discretized  spatial variable $\{r_{t}(\vect{x}); \vect{x} \in \mathcal{L}_r\}$, for $t\in \mathcal{T}_r : \{0,1,\ldots,T,T+1\}$. Define the variable  $\vect{r} =\{\vect{r}_0,\vect{r}_1,\ldots,\vect{r}_T,\vect{r}_{T+1}\} $ and let $\vect{r}_{i:j}$ denote $\{\vect{r}_i,\vect{r}_{i+1},\ldots,\vect{r}_{j}\}, \forall (i,j) \in \mathcal{T}_r^2,i \leq j $. Moreover assume that the temporal $m$-vectors of observations $\vect{d}_t$ for $t \in \mathcal{T}_d : \{0,1,\ldots,T\}$  are available, and define $\vect{d} = \{\vect{d}_0,\vect{d}_1,\ldots,\vect{d}_T\}$ and $\vect{d}_{i:j} = \{\vect{d}_i,\ldots,\vect{d}_j\}$ accordingly. 
The objective of this study is to assess $\vect{r}$ given $\vect{d}$, $[\vect{r}|\vect{d}]$. We define a Kalman type model, represented as a hidden Markov model in a Bayesian inversion framework, in order to retrieve $[\vect{r}|\vect{d}]$. Special attention is given to  assessing the initial state represented by $[\vect{r}_0|\vect{d}]$.
\subsection{Bayesian inversion}
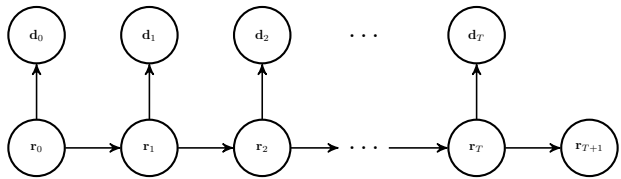
\begin{figure}
	\centering
	\begin{tikzpicture}[->,>=stealth', auto,semithick,node distance=3cm]
	\tikzstyle{every state}=[fill=white,draw=black,thick,text=black,scale=.5, minimum size=1.5cm, inner sep=1]
	\node[state]         (x1)                                                                                         {$\vect{r}_0$};
	\node[state]         (x2)[right of = x1]                                {$\vect{r}_1$};
	\node[state]         (x3)[right of = x2]                                {$\vect{r}_2$};
	\node[draw = none]           (x4)[right of = x3, node distance =1.35cm]                   {$\hdots$}; 
	\node[state]         (x5)[right of =  x4]                               {$\vect{r}_T$};
	\node[state]         (x6)[right of =  x5]                               {$\vect{r}_{T+1}$};
	\edge {x1} {x2}
	\edge {x2} {x3}
	\edge {x3} {x4}
	\edge {x4} {x5}
	\edge {x5} {x6}
	
	\node[state]         (d1)[above of = x1]                              {$\vect{d}_0$};
	\node[state]         (d2)[above of = x2]                              {$\vect{d}_1$};
	\node[state]         (d3)[above of = x3]                              {$\vect{d}_2$};
	\node[draw = none]                           (d4)[right of = d3, node distance = 1.35cm]                  {$\hdots$};
	\node[state]         (d5)[above of = x5]                              {$\vect{d}_T$};
	\edge {x1} {d1}
	\edge {x2} {d2}
	\edge {x3} {d3}
	\edge {x5} {d5}
	
	\end{tikzpicture}
	\caption{Graph of the hidden Markov model}
	\label{HMM}
\end{figure} 
The Kalman type model, phrased as Bayesian inversion,  requires the specification of a prior model for $\vect{r}$ and a likelihood model for $[\vect{d}|\vect{r}]$. The model specified below defines a hidden Markov  model as displayed in Figure \ref{HMM}.
\subsubsection*{Prior model}
The prior model on $\vect{r}$  synthesizes the knowledge  and experience with the spatial variable of interest, and it consists of an initial distribution and a dynamic model:

\paragraph{Initial distribution}
The initial distribution  for the initial state $\vect{r}_0$ is denoted $f(\vect{r}_0)$.
\paragraph{Dynamic model} 
The dynamic model  given the initial state $[\vect{r}_{1:T+1}|\vect{r}_0]$ is defined as,
\begin{align}
f(\vect{r}_{1:T+1}|\vect{r}_0) 
= \prod_{t=0}^T f(\vect{r}_{t+1}|\vect{r}_{t}),
\end{align}
with,
\begin{align*}
[\vect{r}_{t+1}|\vect{r}_t]  = & \omega_t(\vect{r}_t,\vect{\epsilon}_t) \sim f(\vect{r}_{t+1}|\vect{r}_{t}),
\end{align*}
where $\omega_t(\cdot,\cdot)\in \mathbb{R}^n$  is the dynamic function with $\vect{\epsilon}_t$ a random component. Since the dynamic function only involves  the variable at the previous time step $\vect{r}_t$, the  model is a Markov chain.
\subsubsection*{Likelihood model} 
The likelihood  model on $[\vect{d}|\vect{r}]$ provides a link between the variable of interest $\vect{r}$ and the observations $\vect{d}$ and is defined as,
\begin{align}
f(\vect{d}|\vect{r}) = \prod_{t=0}^T f(\vect{d}_t|\vect{r}_t),
\end{align}
with,
\begin{align*}
[\vect{d}_t|\vect{r}_t] =& \nu_t(\vect{r}_t,\vect{\epsilon}_t) \sim f(\vect{d}_t|\vect{r}_t),
\end{align*}
where $\nu_t(\cdot,\cdot) \in \mathbb{R}^m$ is the likelihood function  with $\vect{\epsilon}_t$ a random component.  The likelihood model is defined assuming conditional independence and single state response  and is thus in factored form.
\subsubsection*{Posterior model} 
Bayesian inversion endeavors to assess the posterior distribution of $[\vect{r}|\vect{d}]$,
\begin{align}
\label{full}
f(\vect{r}|\vect{d}) =& \left[\int f(\vect{d}|\vect{r})f(\vect{r})d\vect{r} \right]^{-1} \times f(\vect{d}|\vect{r}) f(\vect{r}) \nonumber \\
= & const \times  f(\vect{d}_0|\vect{r}_0)f(\vect{r}_0) \nonumber \\ \times& \prod_{t=1}^T f(\vect{d}_{t}|\vect{r}_{t})  f(\vect{r}_{t}|\vect{r}_{t-1})f(\vect{r}_{T+1}|\vect{r}_{T}) \nonumber  \\
= &    f(\vect{r}_0|\vect{d})\prod_{t=1}^T f(\vect{r}_{t}|\vect{r}_{t-1},\vect{d}_{t:T})f(\vect{r}_{T+1}|\vect{r}_{T})  
\end{align}
which is a non-stationary  Markov chain for the hidden Markov model with a likelihood model in factored form as defined above, see \cite{moja}. Assessing such a posterior distribution is usually difficult as the normalizing constant is challenging to calculate.

\subsection{Kalman type models}
The current study is limited to Kalman type models. They comprise an initial and a process part.
\paragraph{Initial distribution}
The initial distribution is identical to the initial distribution of the prior model $f(\vect{r}_0)$, and as such captures the initial state of the process. 
Two model classes are later discussed: the Gaussian and the selection-Gaussian classes. 
\paragraph{Process model}
\label{processmodel}
The process model includes the  dynamic component of the prior model and the likelihood  model. It thus characterizes the process dynamics and the observation 
acquisition procedure. The dynamic model is defined by,
\begin{align}
[\vect{r}_{t+1}|\vect{r}_t] =& \mat{A}_t \vect{r}_t + \vect{\epsilon}_{t} \nonumber\\
f(\vect{r}_{t+1}|\vect{r}_t) =& \varphi_n(\vect{r}_{t+1};\mat{A}_t \vect{r}_t,\mat{\Sigma}_{t}^{r|r}),
\end{align}
with forward $(n\times n)$-matrix   $\mat{A}_t $ and  $n$-vector error term $\vect{\epsilon}_{t}$ defined as  centered Gaussian with  covariance $(n\times n)$-matrix $\mat{\Sigma}_{t}^{r|r}$. It defines the dynamic  part of the model, possibly in a transient phase, which is Gauss-linear. 
The likelihood component is defined by,
\begin{align}
[\vect{d}_t|\vect{r}_t] =& \mat{H}\vect{r}_t + \vect{\epsilon}_{t} \nonumber\\ f(\vect{d}_t|\vect{r}_t) =& \varphi_p(\vect{d}_t;\mat{H} \vect{r}_t,\mat{\Sigma}_{t}^{d|r} ),
\end{align}
with the  observation $(m\times n)$-matrix $\mat{H}$ and  the $m$-vector error term $\vect{\epsilon}_{t}$  defined as  centered Gaussian with  covariance $(m\times m)$-matrix $\mat{\Sigma}_{t}^{d|r}$. It represents the observation acquisition procedure which is also Gauss-linear. This process model coincides with the frequently used traditional Kalman model, see \cite{kalman}. 

\subsection{Traditional Kalman model} 
\label{tkf}
The traditional Kalman model is defined by letting the initial distribution  be in the class of Gaussian pdfs,
\begin{align}
\vect{r}_0 \sim f(\vect{r}_0) = \varphi_n(\vect{r}_0;\vect{\mu}_0^r,\mat{\Sigma}_0^r),
\end{align}
with  initial  expectation $n$-vector $\vect{\mu}_0^r$ and   positive definite covariance $(n\times n)$-matrix $\mat{\Sigma}_{0}^{r}$. The Gaussian initial distribution is parametrized by $\Theta^G =(\vect{\mu}_0^r,\mat{\Sigma}_{0}^{r})$. In our spatial study, this initial distribution will be a discretized stationary Gaussian random field. The process model is Gauss-linear and identical to the traditional Kalman type. 

This traditional Kalman model is analytically tractable. The posterior distribution $f(\vect{r}|\vect{d})$ is Gaussian and the posterior distribution parameters can be calculated by algebraic operations  on the parameters of the initial distribution, process model and the observed data. Therefore the assessment of the posterior distribution does not require computationally demanding integrals. The analytical tractability follows from the recursive reproduction of Gaussian pdfs:
\begin{itemize}
	\item  The initial model $f(\vect{r}_0)$ is Gaussian and the likelihood model $f(\vect{d}_0|\vect{r}_0)$ is Gauss-linear, hence the joint model  $f(\vect{r}_0,\vect{d}_0)$ is Gaussian. Consequently, the conditional model $f(\vect{r}_0|\vect{d}_0)$ is Gaussian.
	\item The conditional model  $f(\vect{r}_0|\vect{d}_0)$ is Gaussian and the dynamic model $f(\vect{r}_{1}|\vect{r}_0)$ is Gauss-linear, hence the joint conditional model $f(\vect{r}_1,\vect{r}_0|\vect{d}_0)$ is Gaussian.
\end{itemize} 
By recursion, we obtain that $f(\vect{r}| \vect{d}) = f(\vect{r}_0,\ldots,\vect{r}_{T+1}|\vect{d}_0,\ldots,\vect{d}_{T})$ is Gaussian. 
Note in particular that since $f(\vect{r}|\vect{d})$ is Gaussian, so is  $f(\vect{r}_0|\vect{d})$. This pdf is obtained by marginalization of $f(\vect{r}|\vect{d})$ which, for the Gaussian case, amounts to  removing rows from the expectation vector and rows and columns from the covariance matrix. Additionally, the joint pdf  $f(\vect{r},\vect{d})$ can be  assessed using a simple recursive algorithm, see Algorithm \ref{alg:1} in Appendix \ref{mgf}.

From the joint Gaussian pdf $f(\vect{r},\vect{d})$, the posterior distribution $f(\vect{r}|\vect{d})$ can be analytically assessed. In spatial models, the  grid dimension $n$ may be large while the number of data collection sites $m$ usually is small. The covariance matrix is  a $[n(T+2)+m(T+1)]\times [n(T+2)+m(T+1)]$-matrix and therefore impossible to store for large models. Note that if the aim of the study is clearly defined, as is the case when focus is on $[\vect{r}_{0}|\vect{d}]$, one may only store the model parameters of $[\vect{r}_{0}|\vect{d}]$ where the covariance is a $[n+m(T+1)]\times [n+m(T+1)]$-matrix. Storage is then a lesser issue. 
\subsection{Selection Kalman model}
\label{ekm}
The selection Kalman model is defined by letting the initial distribution be in the class of selection-Gaussian pdfs, see \cite{delpino} and \cite{branco}. This class is defined by considering a pdf from the Gaussian class,
\begin{align*}
f(\tilde{\vect{r}}) =& \varphi_n(\tilde{\vect{r}}; \vect{\mu}_{\tilde{r}},  \mat{\Sigma}_{\tilde{r}})
\end{align*}
with expectation $n$-vector $\vect{\mu}_{\tilde{r}}$ and covariance $(n\times n)$-matrix $\mat{\Sigma}_{\tilde{r}}$. In our spatial study this pdf will represent a discretized stationary Gaussian random field.
Define  further an auxiliary variable $\vect{\nu} \in \mathbb{R}^q$ by a Gauss-linear extension,
\begin{align*}
[\vect{\nu}|\vect{\tilde{r}}]  = \vect{\mu}_{\nu} +\mat{\Gamma}_{\nu|\tilde{r}}(\vect{\tilde{r}}-\vect{\mu}_{\tilde{r}}) + \vect{\epsilon}_{\nu|\tilde{r}}   
\end{align*}
with the expectation $q$-vector $\vect{\mu}_{\nu} $, and the regression $(q\times n)$-matrix  $\mat{\Gamma}_{\nu|\tilde{r}}$ and the centered Gaussian $q$-vector $\vect{\epsilon}_{\nu|\tilde{r}}$, independent of $\vect{\tilde{r}}$, with covariance $(q\times q)$-matrix $\mat{\Sigma}_{\nu|\tilde{r}}$. In the current spatial study the dimension of $\vect{\tilde{r}}$ and $\vect{\nu}$ will be identical. Generally, we have,
\begin{align*}
f(\vect{\nu}|\vect{\tilde{r}}) = \varphi_q(\vect{\nu}; \vect{\mu}_{\nu|\tilde{r}},\mat{\Sigma}_{\nu|\tilde{r}})
\end{align*}
with $\vect{\mu}_{\nu|\tilde{r}}=\vect{\mu}_{\nu} +\mat{\Gamma}_{\nu|\tilde{r}}(\vect{\tilde{r}}-\vect{\mu}_{\tilde{r}})$. As a consequence, $[\vect{\tilde{r}},\vect{\nu}]$ is jointly Gaussian,
\begin{align*}
\begin{bmatrix}\vect{\tilde{r}} \\\vect{\nu} \end{bmatrix}
\sim 
\varphi_{n+q}\left(\begin{bmatrix}\vect{\tilde{r}} \\\vect{\nu} \end{bmatrix}; \begin{bmatrix}
\vect{\mu}_{\tilde{r}} \\
\vect{\mu}_{\nu}
\end{bmatrix}, \begin{bmatrix}
\mat{\Sigma}_{\tilde{r}} & \mat{\Sigma}_{\tilde{r}}\mat{\Gamma}_{\nu|\tilde{r}}^T \\
\mat{\Gamma}_{\nu|\tilde{r}}\mat{\Sigma}_{\tilde{r}} &  \mat{\Sigma}_{\nu}
\end{bmatrix}\right)
\end{align*}
with the covariance $(q \times q)$-matrix $\mat{\Sigma}_{\nu}=\mat{\Gamma}_{\nu|\tilde{r}}\mat{\Sigma}_{\tilde{r}}\mat{\Gamma}_{\nu|\tilde{r}}^T+ \mat{\Sigma}_{\nu|\tilde{r}}$.
Define a selection subset $A \subset \mathbb{R}^q$, and define the class of selection-Gaussian pdfs by $\vect{r}_A=[\vect{\tilde{r}}| \vect{\nu} \in A ]$. In the current spatial study the set $A$ will be separable in $\mathbb{R}^q$. Generally, it follows that,
\begin{align}
f(\vect{r}_A) =& f(\vect{\tilde{r}}| \vect{\nu} \in A)  \\
= & \left[\Phi_q(A;\vect{\mu}_{\nu},\mat{\Sigma}_{\nu}) \right]^{-1} \nonumber\\ \times& \Phi_q(A;\vect{\mu}_{\nu|\tilde{r}}, \vect{\Sigma}_{\nu|\tilde{r}})\times \varphi_n(\vect{\tilde{r}}; \vect{\mu}_{\tilde{r}},\mat{\Sigma}_{\tilde{r}}). \nonumber
\end{align}
This class of pdfs is parametrized by $\vect{\Theta}^{SG} = (\vect{\mu}_{\tilde{r}},\mat{\Sigma}_{\tilde{r}},\vect{\mu}_{\nu},\mat{\Gamma}_{\nu|\tilde{r}},\mat{\Sigma}_{\nu|\tilde{r}},A)$ for all valid parameter sets. The class of selection-Gaussian pdfs is very flexible and may represent multi-modality, skewness and peakedness, see \cite{henning}.
\begin{figure}
	\centering
	\begin{subfigure}{3cm}
		\centering\includegraphics[width=2.5cm]{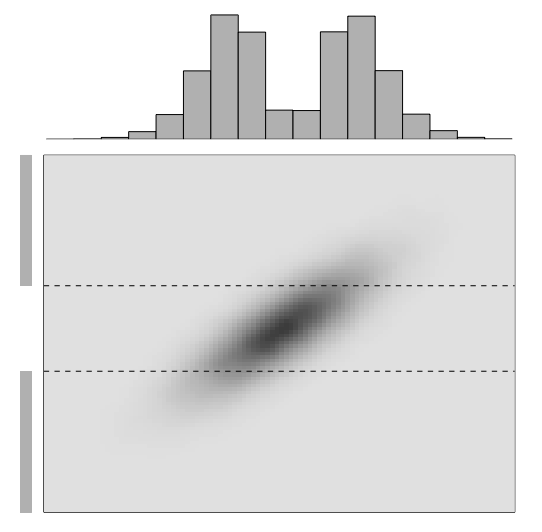}
		\caption{}
		\label{ex:1A}
	\end{subfigure}
	\begin{subfigure}{3cm}
		\centering\includegraphics[width=2.5cm]{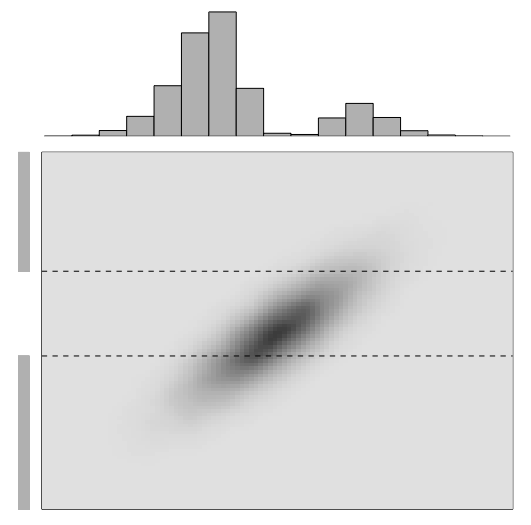}
		\caption{}
		\label{ex:1B}
	\end{subfigure}
	
	\begin{subfigure}{3cm}
		\centering\includegraphics[width=2.5cm]{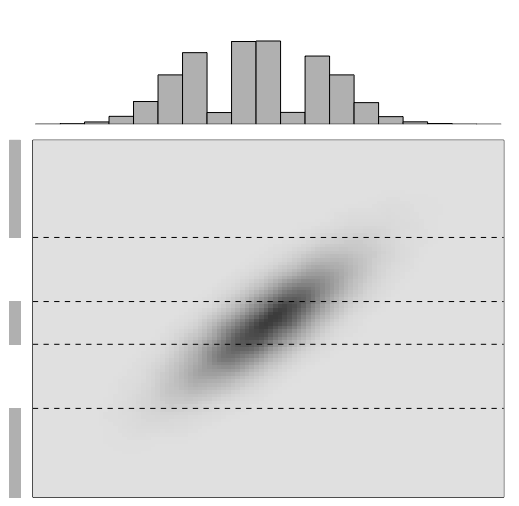}
		\caption{}
		\label{ex:1C}
	\end{subfigure}
	\begin{subfigure}{3cm}
		\centering\includegraphics[width=2.5cm]{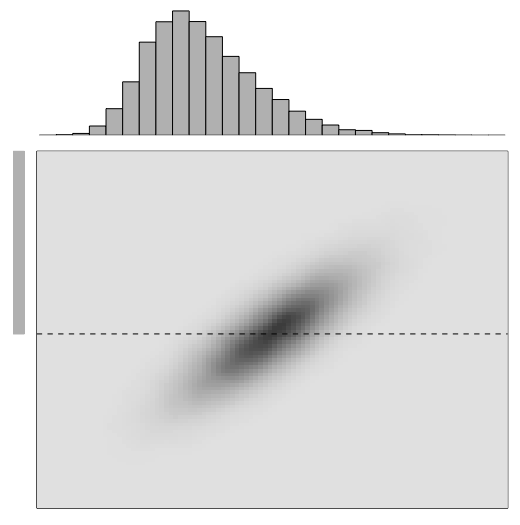}
		\caption{}
		\label{ex:1D}
	\end{subfigure}
	\caption{Realizations of 1D selection-Gaussian pdfs (histogram) with varying selection sets $A \subset \mathbb{R}^n$ (solid gray bars) for a bi-Gaussian pdf $[\tilde{r},\nu]$ (dark gray)}
	\label{ex:1}
\end{figure}
Four one-dimensional selection-Gaussian pdfs are displayed in Figure \ref{ex:1} in order to demonstrate the influence of the selection set $A \subset \mathbb{R}$. The bivariate variable $[r,\nu]$ is bi-Gaussian and identical in all displays, while the selection sets are marked  as solid gray bars along the vertical $\nu$-axis. Figure \ref{ex:1A} contains a selection set  comprised of two segments symmetric about the expectation of $\nu$, making the selection-Gaussian pdf along the horizontal axis bimodal and symmetric. Figure \ref{ex:1B} contains a selection set of two asymmetric segments, making the selection-Gaussian pdf bimodal and asymmetric. Figure \ref{ex:1C} contains a selection set of three segments symmetric about the expectation of $\nu$, making the selection-Gaussian pdf trimodal and symmetric. Lastly, Figure \ref{ex:1D} contains a selection set comprised of only one segment, making the selection-Gaussian pdf skewed. This selection concept can be extended to higher dimensions and even to discretized spatial models.

Note that assigning a null-matrix to $\mat{\Gamma}_{\nu|\tilde{r}}$ entails that $f(\vect{\tilde{r}},\vect{\nu}) = f(\vect{\tilde{r}})f(\vect{\nu})$ and selection on $\vect{\nu}$ does not influence $\vect{\tilde{r}}$. It follows that $f(\vect{{r}}_A) = f(\vect{\tilde{r}})$ is Gaussian. The selection-Gaussian model can therefore be seen as a generalization of the Gaussian one. It can be demonstrated, see \cite{henning}, that the following  recursive reproduction of selection-Gaussian pdfs holds:
\begin{itemize}
	\item  The initial model $f(\vect{r}_0)$ is selection-Gaussian and the likelihood model $f(\vect{d}_0|\vect{r}_0)$ is Gauss-linear, hence the joint model  $f(\vect{r}_0,\vect{d}_0)$ is selection-Gaussian. Moreover, the conditional model $f(\vect{r}_0|\vect{d}_0)$ is selection-Gaussian.
	\item The conditional model  $f(\vect{r}_0|\vect{d}_0)$ is selection-Gaussian and the dynamic model $f(\vect{r}_{1}|\vect{r}_0)$ is Gauss-linear, the joint conditional model $f(\vect{r}_1,\vect{r}_0|\vect{d}_0)$ is therefore selection-Gaussian.
\end{itemize} 
By recursion, we obtain that $f(\vect{r}| \vect{d}) = f(\vect{r}_0,\ldots,\vect{r}_{T+1}|\vect{d}_0,\ldots,\vect{d}_{T})$ is selection-Gaussian. 
Recall that these characteristics are similar to those of the  class  of Gaussian pdfs that makes the traditional Kalman model analytically tractable. The selection Kalman  model is defined with an initial distribution  from the class of selection-Gaussian pdfs and a process model which is Gauss-linear and identical to the traditional Kalman type. From the characteristics of the  class of selection-Gaussian distributions, it follows that the posterior distribution $f(\vect{r}|\vect{d})$ is in the  class of selection-Gaussian distributions and so is $f(\vect{r}_{0}|\vect{d})$.\\
The conditional independence $f(\vect{r}_{t+1},\vect{\nu},\vect{d}_t|\vect{r}_t) =f(\vect{r}_{t+1}|\vect{r}_t)f(\vect{\nu}|\vect{r}_t)f(\vect{d}_t|\vect{r}_t)$ justifies the following algorithm for obtaining $f(\vect{\tilde{r}},\vect{\nu},\vect{\tilde{d}})$ and provide $f(\vect{\tilde{r}},\vect{\nu}|\vect{\tilde{d} = \vect{d}})$.
\begin{alg}{2}{ Joint Selection Kalman Model}
	\begin{itemize} 
		\label{alg:B}
		\item Define 
		\item[]$\vect{\mu}_t^{\tilde{r}} = \mathbb{E}[\vect{\tilde{r}}_{t}]$
		\item[]$\vect{\mu}_0^{\nu} = \mathbb{E}[\vect{\nu}]$
		\item[]$\vect{\mu}_t^{\tilde{d}} =\mathbb{E}[\vect{\tilde{d}}_{t}]$
		\item[]$\mat{\Sigma}^{\tilde{r}\tilde{r}}_{ts} = \Cov(\vect{\tilde{r}}_{t},\vect{\tilde{r}}_s)={\mat{\Sigma}^{\tilde{r}\tilde{r}}_{st}}^T $
		\item[]$\mat{\Sigma}^{\tilde{d}\tilde{d}}_{ts} = \Cov(\vect{\tilde{d}}_{t},\vect{\tilde{d}}_s) ={\mat{\Sigma}^{\tilde{d}\tilde{d}}_{st}}^T $
		\item[]$\mat{\Sigma}^{\nu \nu}_{00} =\Cov(\vect{\nu},\vect{\nu})$
		\item[]$\mat{\Gamma}_{ts}^{\tilde{r}\tilde{d}}=\Cov(\vect{\tilde{r}}_{t},\vect{\tilde{d}}_s)={\mat{\Gamma}_{st}^{\tilde{d}\tilde{r}}}^T$
		\item[]$\mat{\Gamma}^{\tilde{r}\nu}_{t0} = \Cov(\vect{\tilde{r}}_{t},\vect{\nu}) = {\mat{\Gamma}^{\nu\tilde{r}}_{0t}}^T $
		\item[]$\mat{\Gamma}^{\tilde{d}\nu}_{t0} = \Cov(\vect{\tilde{d}}_{t},\vect{\nu}) = {\mat{\Gamma}^{\nu\tilde{d}}_{0t}}^T$
		
		\item Initiate

		\item[]	$\begin{bmatrix}\vect{\tilde{r}}_0 \\\vect{\nu} \end{bmatrix}
		\sim 
		\varphi_{n+q}\left(\begin{bmatrix}\vect{\tilde{r}} \\\vect{\nu} \end{bmatrix}; \begin{bmatrix}
		\vect{\mu}_{\tilde{r}} \\
		\vect{\mu}_{\nu}
		\end{bmatrix}, \begin{bmatrix}
		\mat{\Sigma}_{\tilde{r}} & \mat{\Sigma}_{\tilde{r}}\mat{\Gamma}_{\nu|\tilde{r}}^T \\
		\mat{\Gamma}_{\nu|\tilde{r}}\mat{\Sigma}_{\tilde{r}} &  \mat{\Sigma}_{\nu}
		\end{bmatrix}\right)$
		\item[]$\vect{\mu}_0^{\tilde{r}} = \vect{\mu}_{\tilde{r}}$
		\item[]$\vect{\mu}_0^{\nu} = \vect{\mu}_{\nu} $
		\item[]$\mat{\Sigma}_{00}^{\tilde{r}\tilde{r}}= \mat{\Sigma}_{\tilde{r}}$
		\item[]$\mat{\Sigma}_{00}^{\tilde{r}\nu}= \mat{\Sigma}_{\tilde{r}}\mat{\Gamma}_{\nu|\tilde{r}}^T $
		\item[]$\mat{\Sigma}_{00}^{\nu\nu}= \mat{\Sigma}_{\nu} $
		\item Iterate $t = 0,...,T$
		\begin{itemize}
			\item[]Likelihood model:
			\item[]$\vect{\mu}_t^{\tilde{d}} =\mat{H}\vect{\mu}_t^{\tilde{r}}$
			%\item[]$\vect{\mu}_t^c = \begin{bmatrix}\vect{\mu}_t^u \\ \mat{H}(\vect{\mu}_t^u)_{1:n} \end{bmatrix}$
			\item [] $\mat{\Sigma}_{tt}^{\tilde{d}\tilde{d}}=\mat{H}\mat{\Sigma}^{\tilde{r}\tilde{r}}_{tt}\mat{H}^T+\mat{\Sigma}^{d|r}_t$
			%\item[]$f(\vect{r}_t,...\vect{r}_0,\vect{\nu},\vect{d}_0,...,\vect{d}_{t-1},\vect{d}_t)=N(\vect{\mu}_t^c,\mat{\Sigma}_t^c)$
			\item[]$\mat{\Gamma}_{t0}^{\tilde{d}\nu} =\mat{H} \mat{\Gamma}_{t0}^{\tilde{r}\nu}$
			\begin{itemize}
				\item[] Iterate $s= 0,...,t$
				\begin{itemize}
					\item[]$\mat{\Gamma}_{ts}^{\tilde{r}\tilde{d}}=\mat{\Sigma}^{\tilde{r}\tilde{r}}_{ts}\mat{H}^T$
				\end{itemize}
				\item[] End iterate s
			\end{itemize}
			\begin{itemize}
				\item[]If $t>0$: Iterate $s= 0,...,t-1$
				\begin{itemize}
					\item[]$\mat{\Sigma}_{t,s}^{\tilde{d}\tilde{d}}=\mat{H}\mat{\Gamma}_{ts}^{\tilde{r}\tilde{d}}$
				\end{itemize}
				\item[] End iterate s
			\end{itemize}
			%\item[]$\vect{\mu}_t^c = \begin{bmatrix}\vect{\mu}_t^u \\ \mat{H}(\vect{\mu}_t^u)_{1:n} \end{bmatrix}$
			%\item[]$\mat{\Sigma}_t^c = \begin{bmatrix} \mat{\Sigma}_t^u & \mat{V}^T \\
			%  \mat{V} &  \mat{H}\mat{\Sigma}^{\tilde{r}\tilde{r}}_{tt}\mat{H}^T+\mat{\Sigma}^{d|\tilde{r}}_t\end{bmatrix}$
			\item[]Forwarding model:
			\item[]$\vect{\mu}_{t+1}^{\tilde{r}} = \mat{A}_t\vect{\mu}_{t}^{\tilde{r}}$
			%\item[]$\vect{\mu}_{t+1}^u = \begin{bmatrix}   \mat{A}_t(\vect{\mu}_{t}^c)_{1:n} \\\vect{\mu}_{t}^c \end{bmatrix}$
			\item[] $\mat{\Sigma}_{(t+1)(t+1)}^{\tilde{r}\tilde{r}} = 	\mat{A}_t\mat{\Sigma}^{\tilde{r}\tilde{r}}_{tt}\mat{A}_t^T+\mat{\Sigma}^{r|r}_t$
			%\item[]$f(\vect{r}_{t+1},\vect{r}_t,...\vect{r}_0,\vect{\nu},\vect{d}_0,...,\vect{d}_t)=N(\vect{\mu}_{t+1}^u,\mat{\Sigma}_{t+1}^u)$
			\item[]$\mat{\Gamma}_{t+1,0}^{\tilde{r}\nu} =\mat{A}_t \mat{\Gamma}^{\tilde{r}\nu}_{t0}$
			\begin{itemize}
				\item[] Iterate $s= 0,...,t$
				\begin{itemize}
					\item[]$\mat{\Sigma}^{\tilde{r}\tilde{r}}_{t+1,s} = \mat{A}_t\mat{\Sigma}_{ts}^{\tilde{r}\tilde{r}}$
					\item[]$\mat{\Gamma}_{t+1,s}^{\tilde{r}\tilde{d}}= \mat{A}_t\mat{\Gamma}^{\tilde{r}\tilde{d}}_{ts}$
				\end{itemize}
				\item[] End iterate s
			\end{itemize}
			%\item[]$\vect{\mu}_{t+1}^u = \begin{bmatrix}   \mat{B}(\vect{\mu}_{t}^c)_{1:n} \\\vect{\mu}_{t}^c \end{bmatrix}$
			%\item[]$\mat{\Sigma}_{t+1}^u = \begin{bmatrix}
			%	\mat{B}\mat{\Sigma}^{\tilde{r}\tilde{r}}_{tt}\mat{B}^T+\mat{\Sigma}^{\tilde{r}|\tilde{r}}_t & \mat{W}^T \\
			%	\mat{W} &  \mat{\Sigma}_{c}^t
			%	\end{bmatrix}$
		\end{itemize}
		\item End iterate t
		\begin{align*}
		& f \left( \begin{bmatrix}\vect{\tilde{r}} \\ \vect{\nu} \\\vect{\tilde{d}}\end{bmatrix} \right)=\\
		&\varphi_{n(T+2)+q +m(T+1)}\left(\begin{bmatrix}\vect{\tilde{r}} \\\vect{\nu} \\\vect{\tilde{d}} \end{bmatrix}; \begin{bmatrix}
		\vect{\mu}_{\tilde{r}} \\
		\vect{\mu}_{{\nu}} \\
		\vect{\mu}_{\tilde{d}}
		\end{bmatrix},\begin{bmatrix}
		\mat{\Sigma}^{\tilde{r}\tilde{r}} &  \mat{\Sigma}^{\tilde{r}\nu} & \mat{\Gamma}^{\tilde{r}\tilde{d}} \\
		\mat{\Gamma}^{\nu\tilde{r}} & \mat{\Sigma}^{\nu\nu} & \mat{\Gamma}^{\nu\tilde{d}} \\
		\mat{\Gamma}^{\tilde{d}\tilde{r}} &\mat{\Gamma}^{\tilde{d}\nu} &  \mat{\Sigma}^{\tilde{d}\tilde{d}}
		\end{bmatrix}\right)
		\end{align*}
		is then fully assessed by the algorithm.
		%\item Assess $f(\vect{\tilde{r}}_{0:T+1},\vect{\nu},\vect{\tilde{d}}_{0:T})$
	\end{itemize}
\end{alg}
From the joint Gaussian pdf $f(\vect{\tilde{r}},\vect{\nu},\vect{\tilde{d}})$, the pdf $f(\vect{r}_{A,0}|\vect{d})=f(\vect{\tilde{r}}_0|\vect{\nu}\in A,\vect{\tilde{d}} = \vect{d})$ can  be assessed  by first marginalizing $\vect{\tilde{r}}$ and thereafter sequentially conditioning  on $\tilde{\vect{d}}$ and then on $\vect{\nu}$. The final step, conditioning on  $\vect{\nu} \in A$, is computer demanding even though $\vect{\nu}$ has only dimension $q$.  It is therefore necessary to resort to MCMC sampling to assess the pdf, see \cite{henning} and \cite{genz}.  Algorithm \ref{alg:B} requires that a $[n(T+2)+q+m(T+1)]\times [n(T+2)+q+m(T+1)]$ -matrix be stored, which may be prohibited for large grid size. For targeted studies such as in the one in the following case study where $[\vect{r}_0|\vect{d}]$ is of interest, only a $[n+q+m(T+1)]$ $\times [n+q+m(T+1)]$-matrix needs to be stored.

\label{synthetic_study}
\subsection{Model}
Consider a discretized spatio-temporal continuous random field representing the evolution of a temperature field $\{{r}_t(\vect{x}), \vect{x} \in \mathcal{L}_r\}$, $t \in \mathcal{T}_r : \{0,1,.\ldots, T,T+1\}$; ${r}_t(\vect{x}) \in \mathbb{R}$, as defined in Section \ref{section2}. The number of spatial grid nodes is $n= 21 \times 21$, while temporal reference $T$ is the current time up to $T=50$. The discretized spatial field at time $t$ is represented  by the $n$-vector $\vect{r}_t$.\\
Assume that, given the initial spatial field $\vect{r}_0$, the field evolves according to the advection-diffusion equation, a linear partial differential equation,  
\begin{align*}
\frac{\partial r_t({\vect{x})}}{\partial t} - \lambda \nabla^2 r_t(\vect{x})+ \vect{c}\cdot \nabla   r_t({\vect{x})} =& 0\\
\nabla r_t(\vect{x})\cdot\vect{n} =& 0
\end{align*}
with $\lambda \in \mathbb{R}_+$  the known diffusivity coefficient, $\vect{n}$ the outer normal to the domain and $\vect{c}=[c_1,c_2]$ the known velocity field. 
Define the discretized linear dynamics of the spatial field by,
\begin{align*}
[\vect{r}_{t+1}|\vect{r}_t] =& \mat{A}\vect{r}_t + \vect{\epsilon}_{t}\\
f(\vect{r}_{t+1}|\vect{r}_t) =& \varphi_n(\vect{r}_{t+1};\mat{A} \vect{r}_t,\mat{\Sigma}_{t}^{r|r}) 
\end{align*} 
where the $(n\times n)$-matrix $\mat{A}$ represents the  heat dynamics discretized using finite differences, see Appendix \ref{app:B}, while the centred Gaussian $n$-vector $\vect{\epsilon}_{t}$, with covariance $(n\times n)$-matrix $\mat{\Sigma}_{t}^{r|r} = 0 \times \mat{I}_n$ represents model error. Under these assumptions, the dynamic model is exact which constitutes a limiting case to Gauss-linear models. The spatial variable will then evolve as displayed in Figure \ref{data}.
\begin{table}[!t]
	\caption{Parameter values for the discretized advection-diffusion equation}
	\label{tab:discr}
	\begin{center}
		\begin{tabular}{l|c|r|l|c} % <-- Alignments: 1st column left, 2nd middle and 3rd right, with vertical lines in between
			$\lambda$ & dx & dt & $c_1$ & $c_2$\\
			\hline
			$1.43\times10^{-2}$ & 0.1 & 0.5 & 0 & -0.1
		\end{tabular}
	\end{center}
\end{table}
The observations are acquired in a $m = 5$ location pattern on the spatial grid $\mathcal{L}_r$, see Figure \ref{hm1}, at each temporal node in $\mathcal{T}_d$, providing the set of $m$-vectors $\{\vect{d}_t$, $t \in \mathcal{T}_d\}$. The corresponding likelihood model is defined as,
\begin{align*}
[\vect{d}_t|\vect{r}_t] =& \mat{H}\vect{r}_t + \vect{\epsilon}_{t}\\
f(\vect{d}_{t}|\vect{r}_t) =& \varphi_{m}(\vect{d}_t;\mat{H} \vect{r}_t,\mat{\Sigma}^{d|r}_t) 
\end{align*}
where the observation $(m\times n)$-matrix $\mat{H}$ is a binary selection matrix, see Appendix \ref{app:B}, while the centered Gaussian $m$-vector $\vect{\epsilon}_{t}$ with covariance $(m\times m)$-matrix  $\mat{\Sigma}^{d|r}_t =\sigma_{d|r}^2 \times \mat{I}_m$ with $\sigma_{d|r} = 0.1$, represents independent observation errors. Under these assumptions, the likelihood model is Gauss-linear. The observations in the synthetic case are displayed as time series in Figure \ref{data2}.\\
\begin{figure}[!t]
	\centering
	\includegraphics[width=0.5\textwidth]{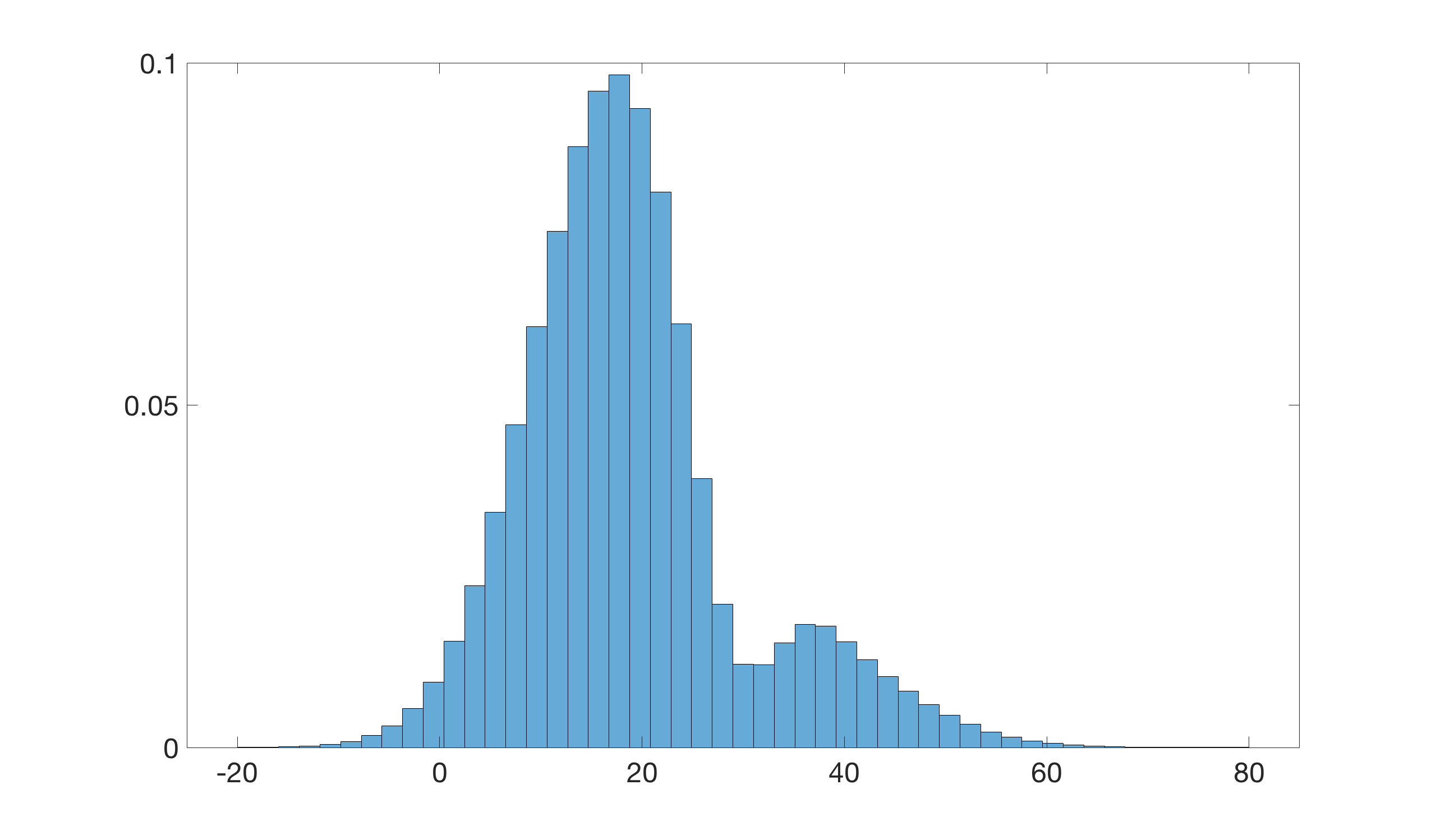}
	\caption{Typical marginal distribution of the initial model}
	\label{priormarg1}
\end{figure}
The prior beliefs of the initial state $\vect{r}_0$ is spatially stationary since the location of the extreme event is unknown. The beliefs on the marginal values are however bi-modal either at the normal level or, less likely, at the high  event value, the marginal pdf should therefore be as sketched in Figure \ref{priormarg1}. The traditional Kalman model requires the initial distribution to be a Gaussian model, with Gaussian marginal pdfs, which cannot capture bimodality.
\begin{figure}[!t]
	\centering
	\includegraphics[width=\textwidth]{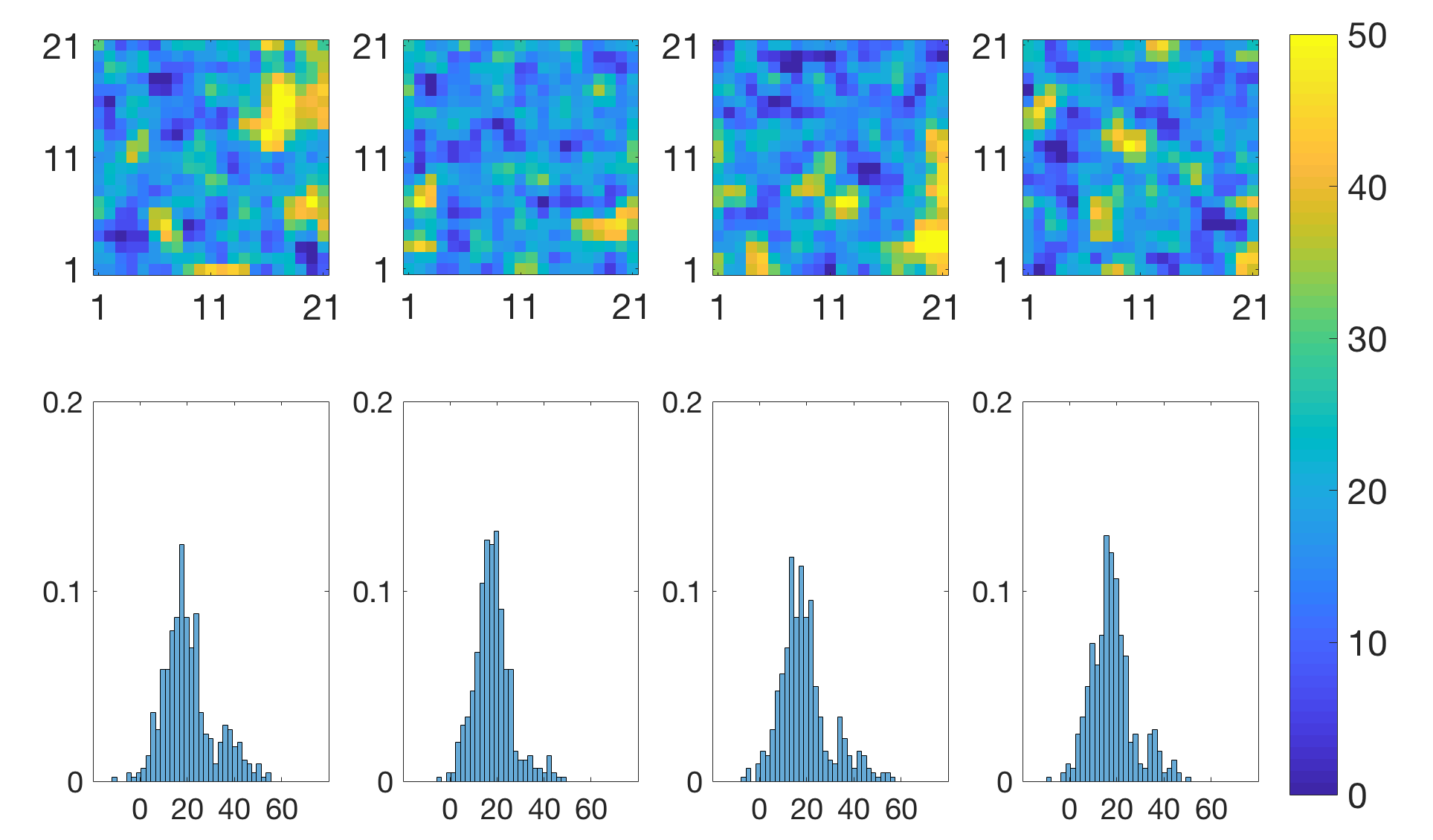}
	\caption{Realizations from the initial selection-Gaussian model; maps (upper), spatial histograms (lower)}
	\label{initialreal1}
\end{figure}
\begin{figure}[!t]
	\centering
	\includegraphics[width=\textwidth]{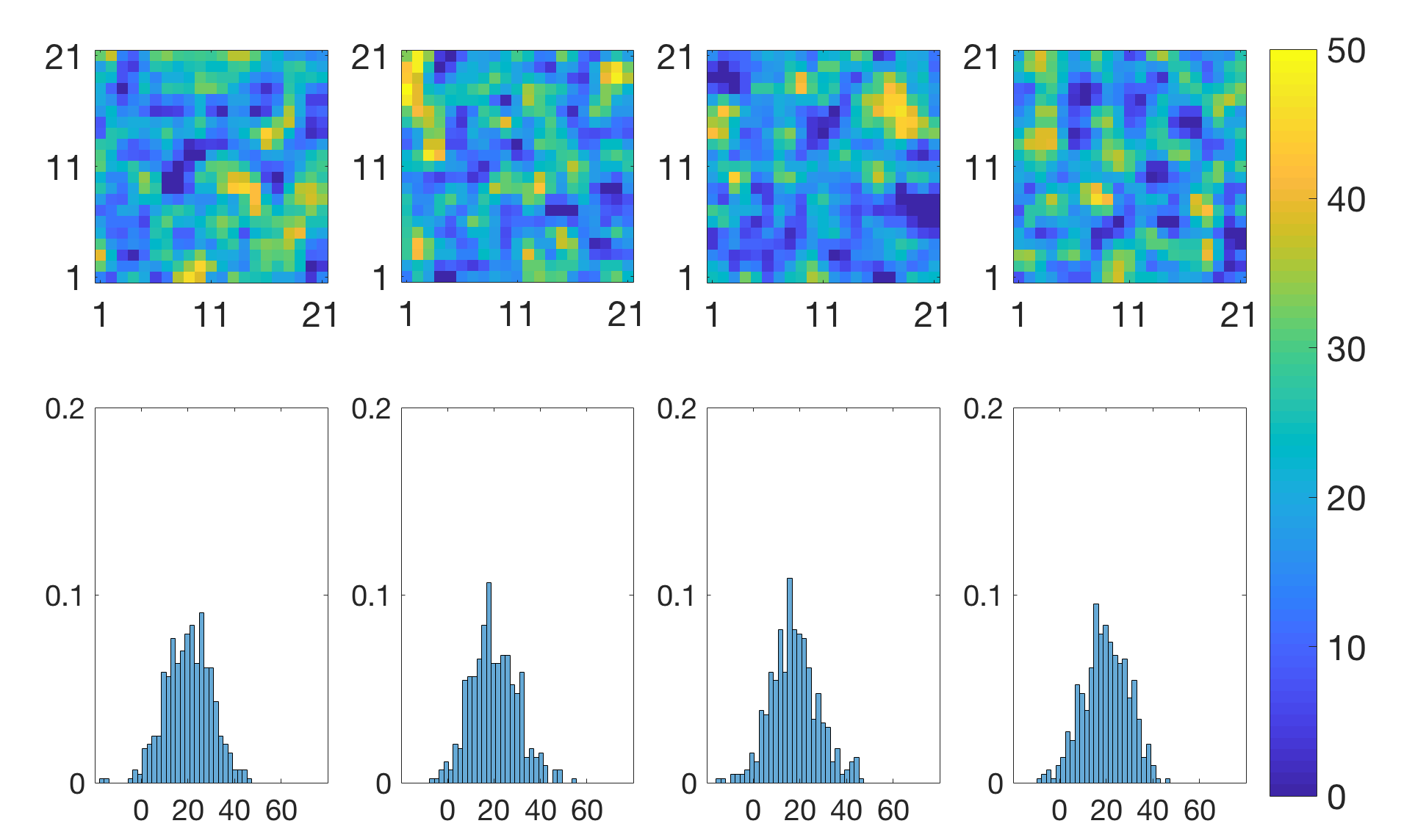}
	\caption{Realizations from the initial Gaussian model; maps (upper), spatial histograms (lower)}
	\label{initialreal2}
\end{figure}
\begin{figure}[!t]
	\centering
	\includegraphics[width=\textwidth]{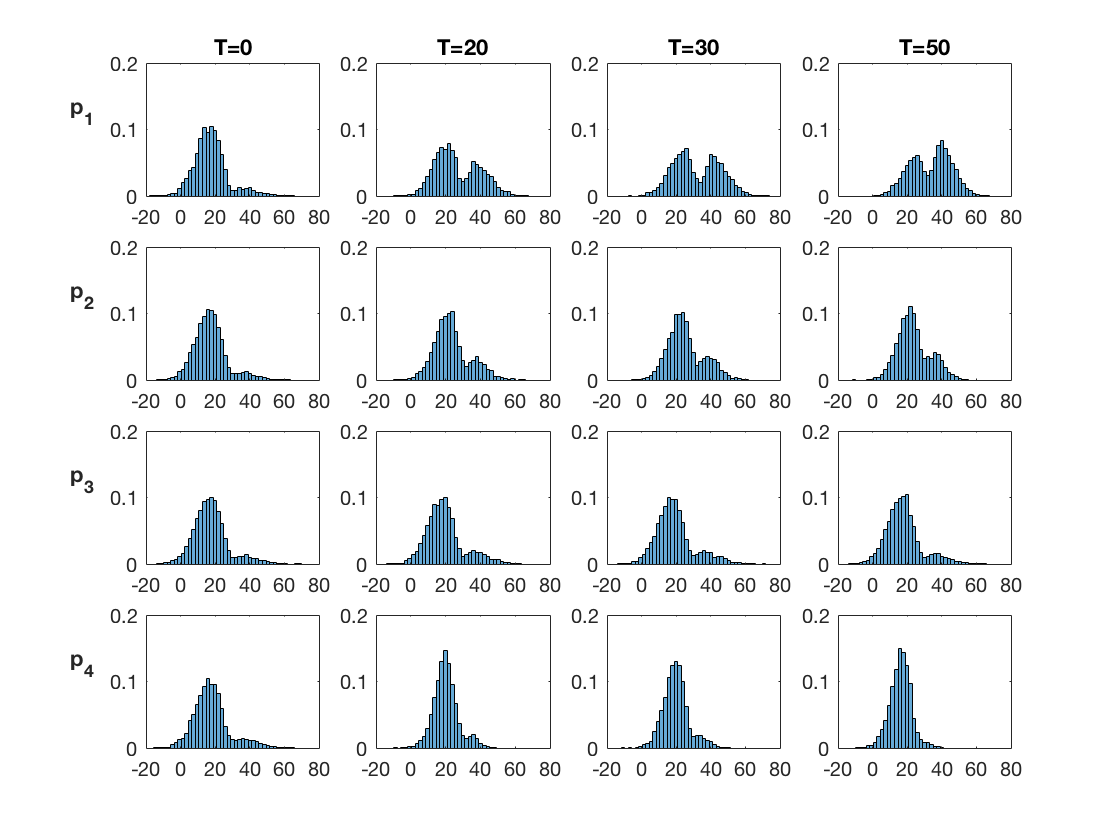}
	\caption{Marginal pdfs at monitoring locations for increasing current time $T$ from the  inversion with the selection Kalman model}
	\label{ressg}
\end{figure}
\begin{figure}[!t]
	\centering
	\includegraphics[width=\textwidth]{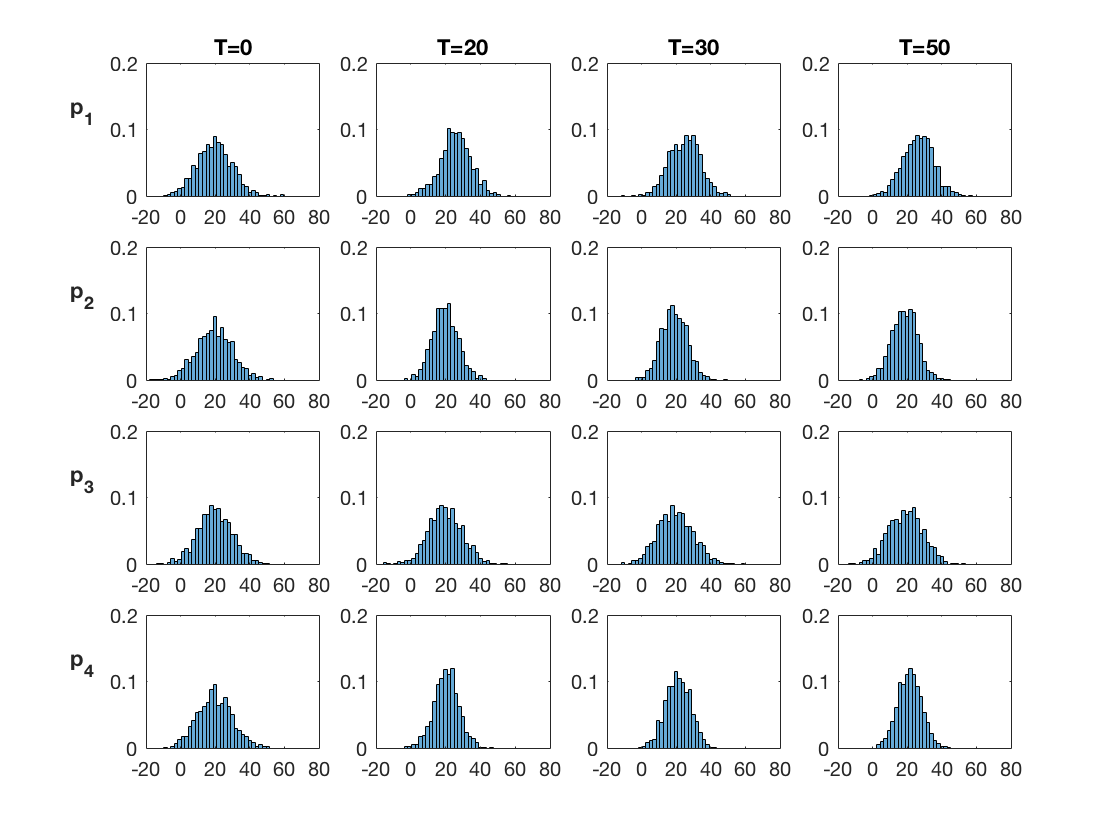}
	\caption{Marginal pdfs at monitoring locations for increasing current time $T$  from the  inversion with the traditional Kalman model}
	\label{resg}
\end{figure}
\begin{figure*}[!t]
	\centering
	\includegraphics[width=\textwidth]{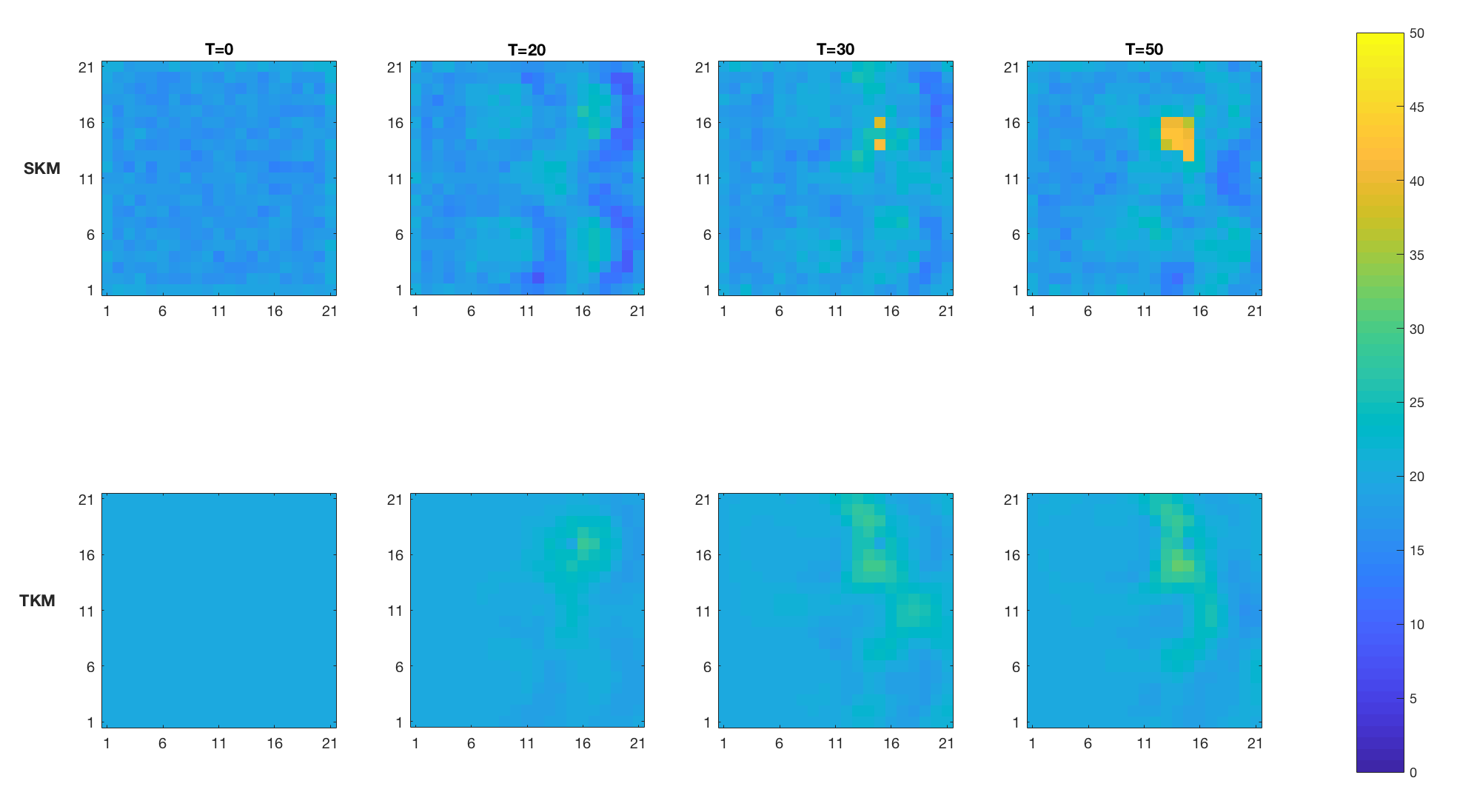}
	\caption{MMAP predictions of the initial state for increasing current time $T$ from the  inversion with the selection Kalman model (upper) and with the traditional Kalman model (lower)}
	\label{MMAP}
\end{figure*}
\begin{figure}[!t]
	\begin{subfigure}{0.5\textwidth}
		\includegraphics[width=\linewidth]{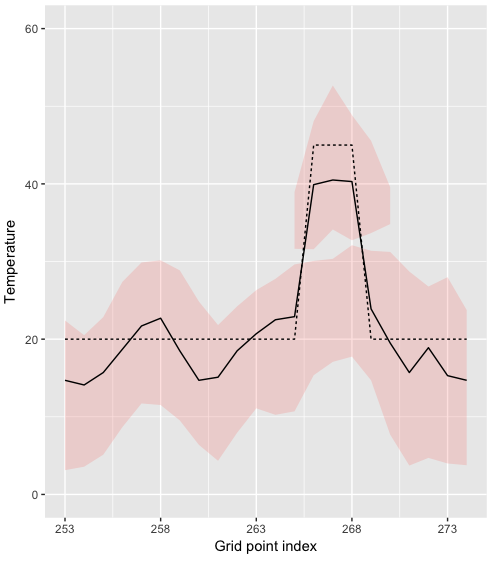}
		\label{HDI}
	\end{subfigure}
	\hspace*{\fill} % separation between the subfigures
	\begin{subfigure}{0.5\textwidth}
		\includegraphics[width=\linewidth]{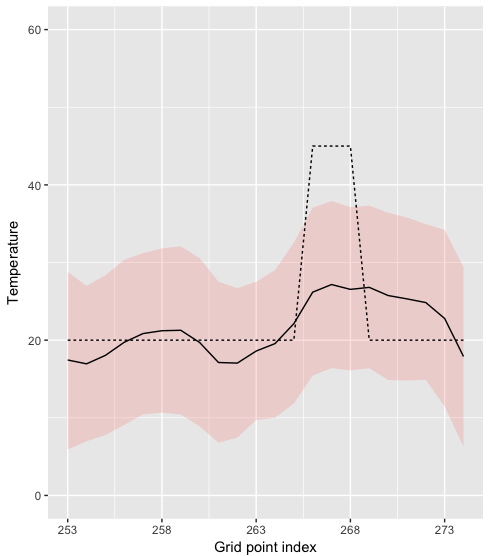}
		\label{CI}
	\end{subfigure}
	\hspace*{\fill} % separation between the subfigures
	\caption{MMAP predictions (solid black line) with HDI 0.8 (red) intervals in cross section A-A' of initial state at current time $T=50$ with selection Kalman model (left) and with traditional Kalman model (right). True cross section (dotted line).} \label{fig:2} 
\end{figure}	
\begin{figure}[!t]
	\centering
	\includegraphics[width=\textwidth]{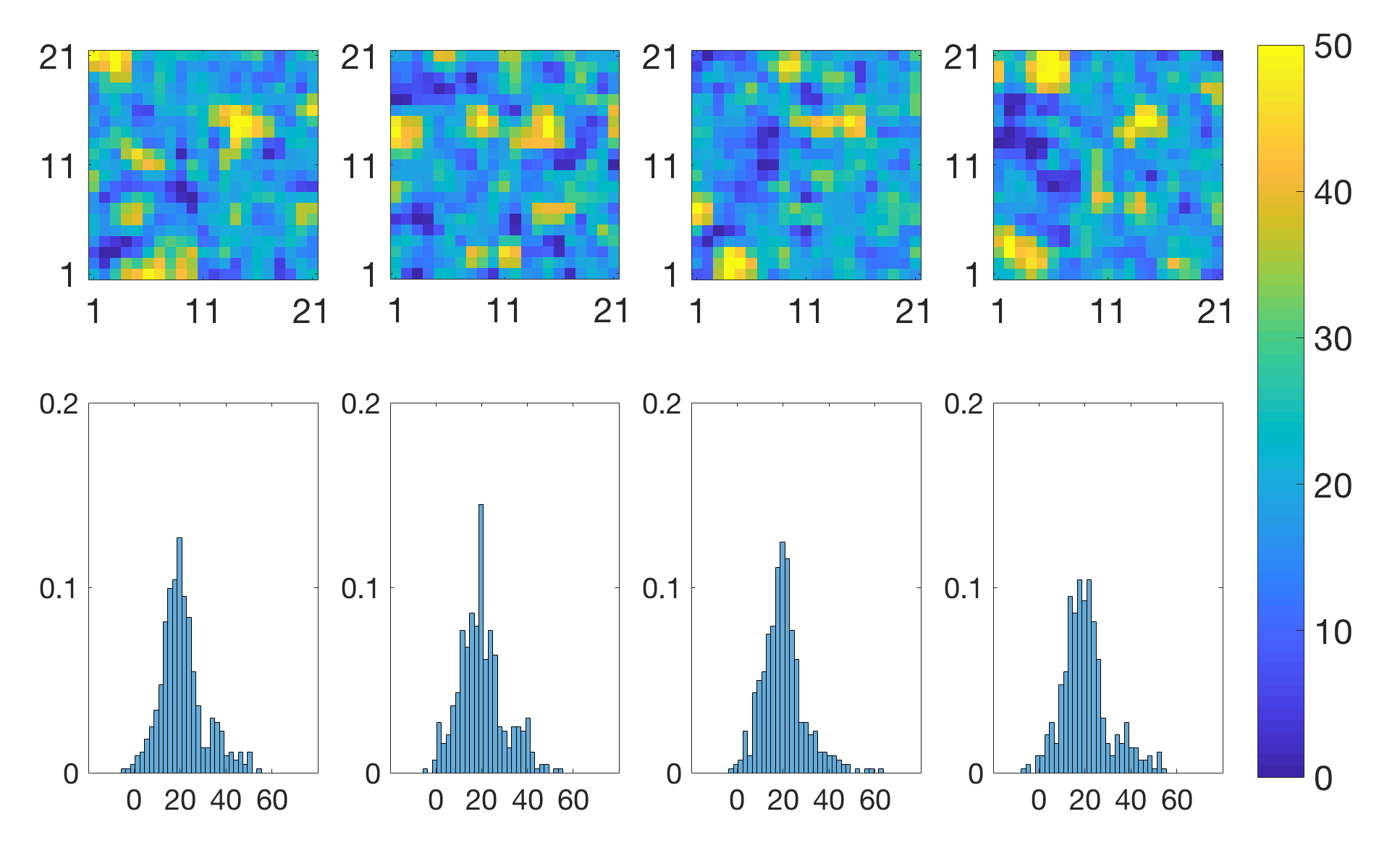}
	\caption{Realizations of the initial state at current time $T=50$ from the inversion with the selection Kalman model}
	\label{finalreal1}
\end{figure}
\begin{figure}[!t]
	\centering
	\includegraphics[width=\textwidth]{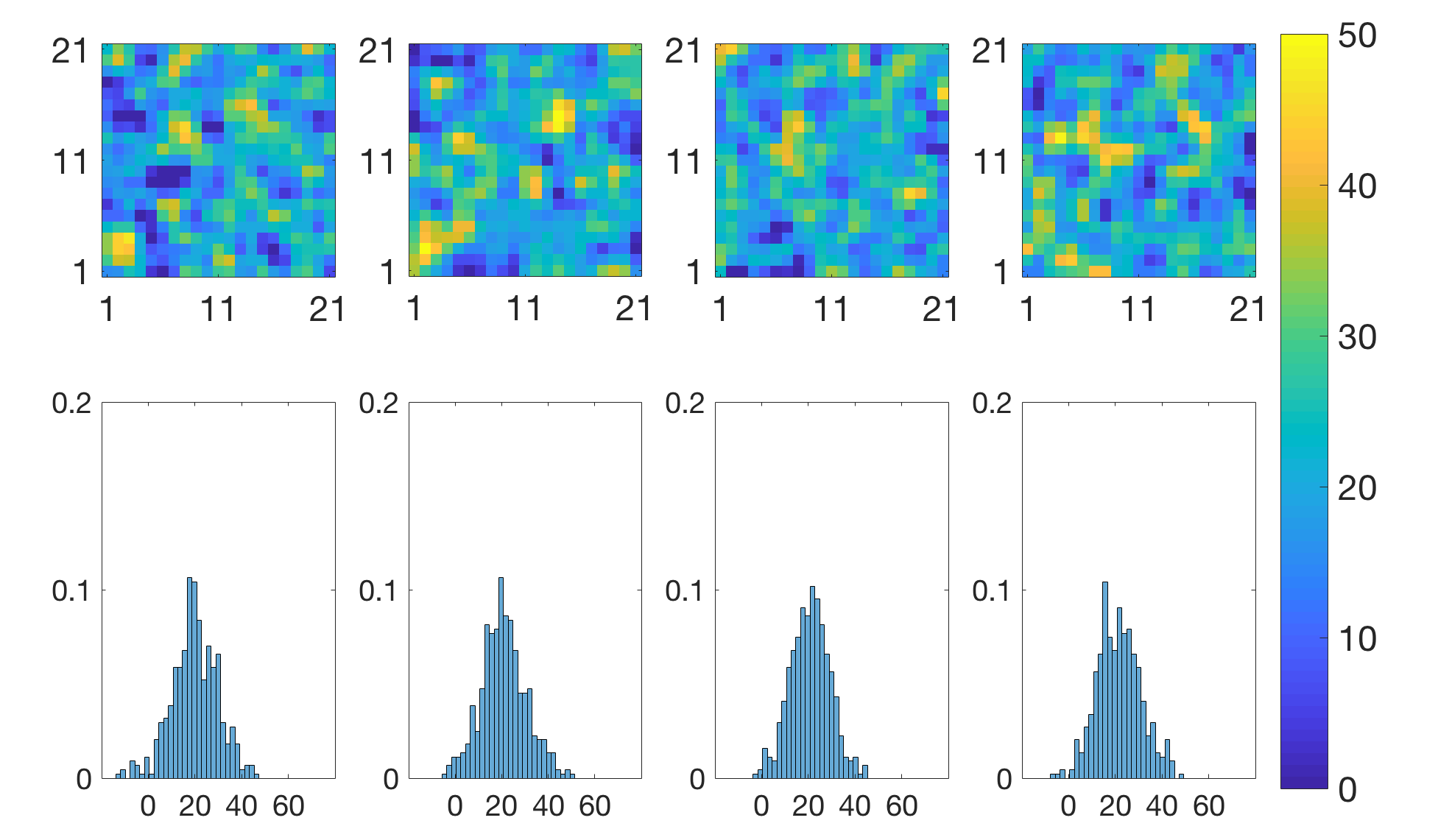}
	\caption{Realizations of the initial state at current time $T=50$ from the inversion with the  traditional Kalman model}
	\label{finalreal2}
\end{figure}

The initial spatial distribution in the selection Kalman model is defined to be in the class of  selection-Gaussian models which can capture bimodality, see  \cite{henning}. The selection-Gaussian model is defined considering a discretized stationary Gaussian random field, 
\begin{align*}
f(\tilde{\vect{r}}) = \varphi_n(\vect{\tilde{r}};\mu_{\tilde{r}}
\vect{i}_{n},\sigma_{\tilde{r}}^2\mat{\Sigma}_{\tilde{r}}^{\rho})
\end{align*}
with expectation and variance levels, $\mu_{\tilde{r}}$ and $\sigma_{\tilde{r}}^2$ respectively. The spatial correlation ($n\times n$)-matrix $\mat{\Sigma}_{\tilde{r}}^{\rho}$ is defined by an isotropic second order exponential spatial correlation function $\rho_{\tilde{r}}(\tau) = \exp{(-\tau^2/\delta^2)}; \tau \in \mathbb{R}_{+}$. Define the auxiliary variable $\vect{\nu} \in \mathbb{R}^n$ given $\vect{\tilde{r}}$,
\begin{align*}
[\vect{\nu}|\vect{\tilde{r}}] =& \gamma(\vect{\tilde{r}} - \mu_{\tilde{r}}
\vect{i}_{n}) + \vect{\epsilon} \\
f(\vect{\nu}|\vect{\tilde{r}}) =& \varphi_n(\vect{\nu}; \gamma  (\vect{\tilde{r}} - \mu_{\tilde{r}} \vect{i}_{n}), (1-\gamma^2)\mat{I}_n)\\
=& \prod_{i=1}^{n}   \varphi_1(\nu_i; \gamma  ({\tilde{r}}_i - \mu_{\tilde{r}} ), (1-\gamma^2))
\end{align*}
with coupling parameter $\gamma \in \mathbb{R}_{[-1,1]}$ and centered Gaussian independent  $n$-vector $\vect{\epsilon}$ with variance $(1-\gamma^2)$. Note that this pdf is in factored form. Consequently the joint pdf of $[\vect{\tilde{r}},\vect{\nu}]$ is,
\begin{equation*}
\begin{bmatrix}\vect{\tilde{r}} \\\vect{\nu} \end{bmatrix}
\sim 
\varphi_{2n}\left(\begin{bmatrix}\vect{\tilde{r}} \\\vect{\nu} \end{bmatrix}; \begin{bmatrix}
\mu_{\tilde{r}}\vect{i}_{n}\\
0\vect{i}_n
\end{bmatrix}, \begin{bmatrix}
\sigma_{\tilde{r}}^2\mat{\Sigma}_{\tilde{r}}^{\rho} & \sigma_{\tilde{r}}^2\gamma \mat{\Sigma}_{\tilde{r}}^{\rho} \\
\sigma_{\tilde{r}}^2\gamma\mat{\Sigma}_{\tilde{r}}^{\rho} &  \sigma_{\tilde{r}}^2\gamma^2\mat{\Sigma}_{\tilde{r}}^{\rho} + (1-\gamma^2) \mat{I}_n
\end{bmatrix}\right).
\end{equation*}
Define a separable selection set $\vect{A} \in \mathbb{R}^n$ such that $\vect{A} = \bigcup A_i, A_i=A_j; (i,j) \in \{1,\ldots,n\}$, and define the selection Gaussian random field $\vect{r}_A$ as, 
\begin{align*}
\vect{r}_A =& [\vect{\tilde{r}}|\vect{\nu}\in \vect{A}]\\
f(\vect{r}_A) =& \left[{\Phi_n(\vect{A},0 \vect{i}_{n},\sigma_{\tilde{r}}^2\gamma^2\mat{\Sigma}_{\tilde{r}}^{\rho} + (1-\gamma^2) \mat{I}_n)}\right]^{-1}\\ 
\times&\prod_{i=1}^{n}   \Phi_1(A_i; \gamma  ({\tilde{r}}_i - \mu_{\tilde{r}} ), (1-\gamma^2))\\\times& \varphi_n(\vect{r}_A;\mu_r
\mat{i}_{n},\sigma_{\tilde{r}}^2\mat{\Sigma}_{\tilde{r}}^{\rho})
\end{align*}
The initial distribution for $\vect{r}_0$ is defined as $f(\vect{r}_{A,0})$ with  parameter values as listed in Table \ref{tab:sgpar}. Note that after selection on the auxiliary variable $\vect{\nu}$ is made, the expectation and variance of the resulting $\vect{r}_A$ will no longer be $\mu_{\tilde{r}}
\vect{i}_{n}$ and $\sigma_{\tilde{r}}^2\mat{\Sigma}_{\tilde{r}}^{\rho}$. Figure \ref{initialreal1} display four realizations with spatial histograms from the initial distribution. The initial distribution is spatially stationary, except for boundary effects, while the marginal pdfs are bi-modal. Hence this distribution captures the possibility for some high-valued events as displayed in Figure \ref{priormarg1}.

The initial distribution in the alternative model that constitutes the traditional Kalman model must be a Gaussian pdf,
\begin{align*}
f(\vect{r}) = \varphi_n(\vect{r};\mu_r
\mat{i}_{n},\sigma_{r}^2\mat{\Sigma}_{r}^{\rho})
\end{align*}
with expectation and variance levels, $\mu_r$ and $\sigma_r^2$, respectively and spatial correlation ($n\times n$)-matrix $\mat{\Sigma}_r^{\rho}$ defined by a second order spatial correlation function $\rho_{r}(\tau) = \exp{(-\tau^2/\delta^2)};\tau \in \mathbb{R}_{+}$. The alternative  initial distribution for $\vect{r}_0$ is defined as $f(\vect{r}_0)$ with  parameter values  listed in Table \ref{tab:gpar}.
\begin{table}[!t]
	\caption{Parameters for the selection-Gauss initial model}
	\label{tab:sgpar}
	\begin{center}
		\begin{tabular}{c|c|c|c|c} % <-- Alignments: 1st column left, 2nd middle and 3rd right, with vertical lines in between
			$\mu_{\tilde{r}}$ & $\sigma_{\tilde{r}}$ & $\delta$ &  $\gamma$ &  A \\
			\hline
			28.75 & 10 & 0.15  & 0.95  & $(]-\infty,-0.2]\cap[0.5,+\infty[)^n$
		\end{tabular}
	\end{center}
\end{table}
\begin{table}[!t]
	\caption{Parameter values for the  Gaussian initial model}
	\label{tab:gpar}
	\begin{center}
		\begin{tabular}{c|c|c} % <-- Alignments: 1st column left, 2nd middle and 3rd right, with vertical lines in between
			$\mu_{r}$ & $\sigma_{r}$ & $\delta$  \\
			\hline
			20 & 10 & 0.15
		\end{tabular}
	\end{center}
\end{table}
Figure \ref{initialreal2} displays four realizations with associated spatial histograms from the alternative initial Gaussian distribution. This figure can be compared to Figure \ref{initialreal1} for the initial selection-Gaussian distribution, and one observes that both distributions are spatially stationary, but only the selection-Gaussian distribution can capture bi-modality in the marginal pdf.
In the next section, we demonstrate the effect of specifying different initial distributions on the identification and characterization of extreme events occurring at $t=0$. 
\subsection{Results}
Consider the initial spatial variable at $t=0$ as displayed in Figure \ref{hm1} with $\vect{r}_0$ taking value $20 $ everywhere on the grid $\mathcal{L}_r$ except in a nine node square in the upper right quadrant where the value is $45 $. This square area is termed the extreme event. The five observation  locations are also displayed in the figure. Note further that none of these locations are inside the event. Figure \ref{data} displays the temporal evolution of the spatial variable $\vect{r}_t$ at $t=0,20,30,50$. Note  that the field is in a transient phase from injection of the event  at $t=0$ towards equilibrium. Moreover, the field drifts downwards. In  Figure \ref{data2}, the actual observations $\vect{d} = \{\vect{d}_0,\ldots,\vect{d}_T\}$ are presented. Note that at current time $T=0$ all observations are close to $20 $, it is only later on that the effects of the diffusion  of the  event are observed at some of the  observation locations.
The challenge is to restore $\vect{r}_0$ based on the observations $\vect{d} = \{\vect{d}_0,\ldots,\vect{d}_T\}$ and to do so reliably for a current time $T$ as small as possible. We use two alternative models, the selection Kalman model and the traditional Kalman model, to make this $\vect{r}_0$ restoration, and compare the results. 
Both the  selection and traditional  Kalman models have been fully specified in the previous section. Moreover the algorithms used to assess the inversion challenge are defined. Consequently, the posterior distributions   $f(\vect{r}_{A,0}|\vect{d}_{0:T})$ and  $f(\vect{r}_0|\vect{d}_{0:T})$ for the selection and traditional  Kalman models  respectively are analytically tractable.  The former is a selection-Gaussian pdf while the latter is a Gaussian pdf. We compare the two posterior distributions for increasing current time $T$, and evaluate their respective ability to restore $\vect{r}_0$ as displayed in  Figure \ref{hm1}.
In order to evaluate the results, we present various characteristics of the posterior distributions for increasing current time $T$: 
\begin{enumerate}
	\item Marginal pdfs at four monitoring locations as displayed in Figure \ref{hm1},
	\begin{align}
	f(r_{A,0,i}|\vect{d}_{0:T})=\int f(\vect{r}_{A,0}|\vect{d}_{0:T}) d\vect{r}_{A,0,-i} \quad i=1,\ldots,4
	\end{align}
	and similarly  for $f(r_{0,i}|\vect{d}_{0:T})$ based on $f(\vect{r}_{0}|\vect{d}_{0:T})$.
	\item Spatial prediction based on a marginal maximum a posteriori (MMAP) criterion,
	\begin{align}
	\vect{\hat{r}}_{A,0} =& \MMAP\{ \vect{r}_{A,0}|\vect{d}_{0:T}\}\\
	=& \{ \MAP\{r_{A,0,i}|\vect{d}_{0:T}\};i = 1,2,\ldots,n\} \nonumber\\
	=& \{\argmax \{f(r_{A,0,i}|\vect{d}_{0:T})\}, i=1,2,\ldots,n\} \nonumber
	\end{align}
	and similarly for $\vect{\hat{r}}_{0}$ based on $f(\vect{r}_{0}|\vect{d}_{0:T})$. This MMAP criterion is used as the marginal posterior model may be multi-modal. For uni-modal symmetric posterior distributions such as  the Gaussian one, the MMAP predictor coincides with the expectation predictor. 
	\item The MMAP prediction and the associated  $0.80$ prediction interval along a horizontal profile A-A', see Figure \ref{hm1} . The prediction interval is computed as the highest density interval (HDI), see \cite{hdi}, which entails that the prediction intervals may consist of several intervals for multimodal posterior pdfs.
	\item Realizations from the posterior pdfs $f(\vect{r}_{A,0}|\vect{d}_{0:T})$ and $f(\vect{r}_{0}|\vect{d}_{0:T})$.
	%\item Analysis of the root mean squared error (RMSE) and of the mean absolute error (MAE).
\end{enumerate}
The posterior distribution $f(\vect{r}_{A,0}|\vect{d}_{0:T})$ is a selection-Gaussian pdf and all marginal pdfs are selection-Gaussian and analytically assessable. Due to the coupling to the auxiliary variable, the marginal pdfs and spatial predictions are most efficiently obtained via simulation based inference using a Metropolis Hastings block sampling algorithm, see \cite{henning}. The posterior model $f(\vect{r}_{0}|\vect{d}_{0:T})$ from the traditional Kalman model is a Gaussian pdf and all marginal pdfs are Gaussian and analytically tractable. Therefore, the marginal pdfs and spatial predictions  can easily be obtained.
\\
Figure \ref{ressg} displays the marginal posterior pdfs based  on the selection Kalman model at the four monitoring locations, vertically, for increasing current time $T$, horizontally. At current time $T=0$, all pdfs are virtually identical to the marginal pdf of the stationary initial model. As current time $T$ increases, and the observations are assimilated, one observes substantial differences between the marginal pdfs at the monitoring locations. The height of the high-value  mode increases depending on the proximity of monitoring location to the event, as expected. The posterior marginal pdf at observation location 1 clearly indicates that it lies in the event already at current time $T=20$ as the high-value mode is increasing. At location 2 the high-value mode also increases somewhat at $T=20$, but does not increase more thereafter. This monitoring location is outside the event, although fairly close to it. Location 3 is far from both the event and observation locations and the posterior marginal pdf remains almost identical to the prior model. Lastly location 4 is far from the event but close to an observation location at which the observations remain stationary, hence the low-value mode  grows to be completely dominant. 
\\
Figure \ref{resg} displays the  marginal pdfs from the traditional Kalman model. These marginal posterior pdfs are also virtually identical at current time $T=0$. As current time $T$ increases the marginal pdfs at the monitoring locations are indeed different as they are shifting. However, this shift is difficult to observe. By using the selection Kalman model, the indications of an event  in the correct location can be observed from current time $T=20$, while one can hardly observe any indications of it if the traditional Kalman model is used. 
\\
The upper panels of Figure \ref{MMAP} display the MMAP spatial prediction based on the selection Kalman model for increasing current time $T$. At current time $T=0$, the predictions are virtually constant bar some boundary effect as the initial  prior model is stationary. As current time $T$ increases, indications of the high-value  event appear at $T=30$, it is however  at $T=50$ that correct location and spatial extent are identified. The prediction value of the event is  very close to the correct value of $45 $. The background value  is predicted  with some variability around the expected $20 $. The lower panels of Figure \ref{MMAP} present the corresponding spatial predictions based on the traditional Kalman model.  As current time $T$ increases, indications of something occurring in the  event area appears, but the location is uncertain and the spatial extent only vaguely outlined. Moreover the predicted value in the event area is  much lower than the correct value $45 $. The background value is however fairly precisely predicted around the expected $20 $.
The circular features centered about the observation locations that appear on the predictions based on the selection Kalman model in Figure \ref{MMAP} are not artifacts. These features are also present on the predictions based on the traditional Kalman model, although less prominent.
\begin{table}[!t]
	\caption{RMSE of the predictors based on the selection Kalman model (SKM) and the traditional Kalman model (TKM).}
	\label{tab:RMSE}
	\begin{center}
		\begin{tabular}{c|c c c c} % <-- Alignments: 1st column left, 2nd middle and 3rd right, with vertical lines in between
			& $T=0$ & $T=20$ & $T=30$ & $T=50$\\
			\hline
			SKM & 31.5 & 3.92 & 4.17 & 2.76 \\
			TKM & 3.61 & 3.53 & 3.48 & 3.33
		\end{tabular}
	\end{center}
\end{table}
\\
The root mean square error (RMSE) criterion is used to quantify  the difference between the MMAP predictions in Figure \ref{MMAP}  and the truth in Figure \ref{hm1}.  This criterion favors smooth Gaussian models. Table \ref{tab:RMSE} displays the RMSE  values of the two models for increasing time $T$. We observe mostly a decreasing RMSE as T increases. Initially the traditional Kalman model  prediction has by far the smallest RMSE but as the observations are assimilated, the two predictions appear more and more alike. Lastly, at $T=50$, the selection Kalman model prediction adapts better to the truth since it can represent extreme events. 
\\
Figure \ref{fig:2} displays the MMAP predictions with associated  $0.80$ prediction intervals along the horizontal profile A-A'. The prediction  from the selection Kalman model captures the event while the prediction from the traditional Kalman model  barely indicates the event. The prediction intervals follow the same pattern. Note, however, that the prediction intervals of the selection Kalman model may appear as two intervals close to the event since the marginal posterior models are bimodal. By using the selection Kalman model, the location, spatial extent and value of the extreme event is very precisely predicted at current time $T=50$. Predictions based on the traditional Kalman model are less precise and rather blurred. 
\begin{figure}[!t]
	\centering
	\includegraphics[width=0.6\textwidth]{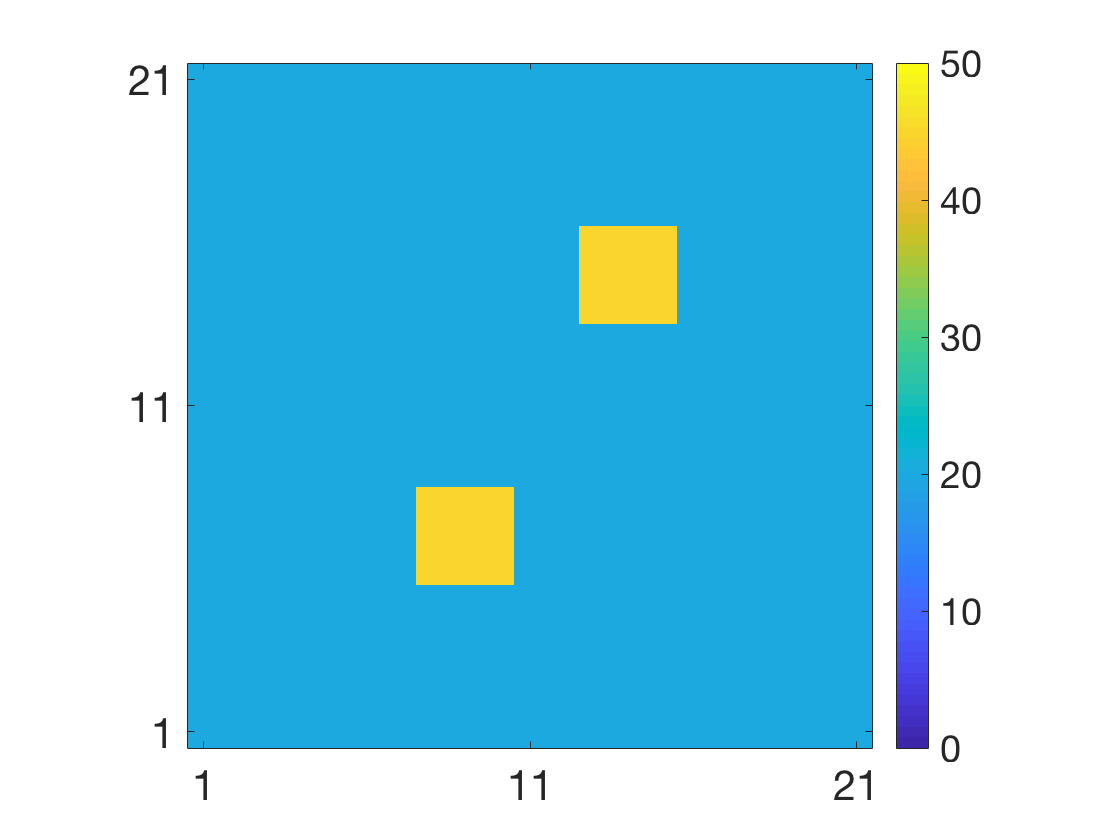}
	\caption{ Initial state for the two-event test case}
	\label{supp1}
\end{figure}
\begin{figure}[!t]
	\centering
	\includegraphics[width=0.6\textwidth]{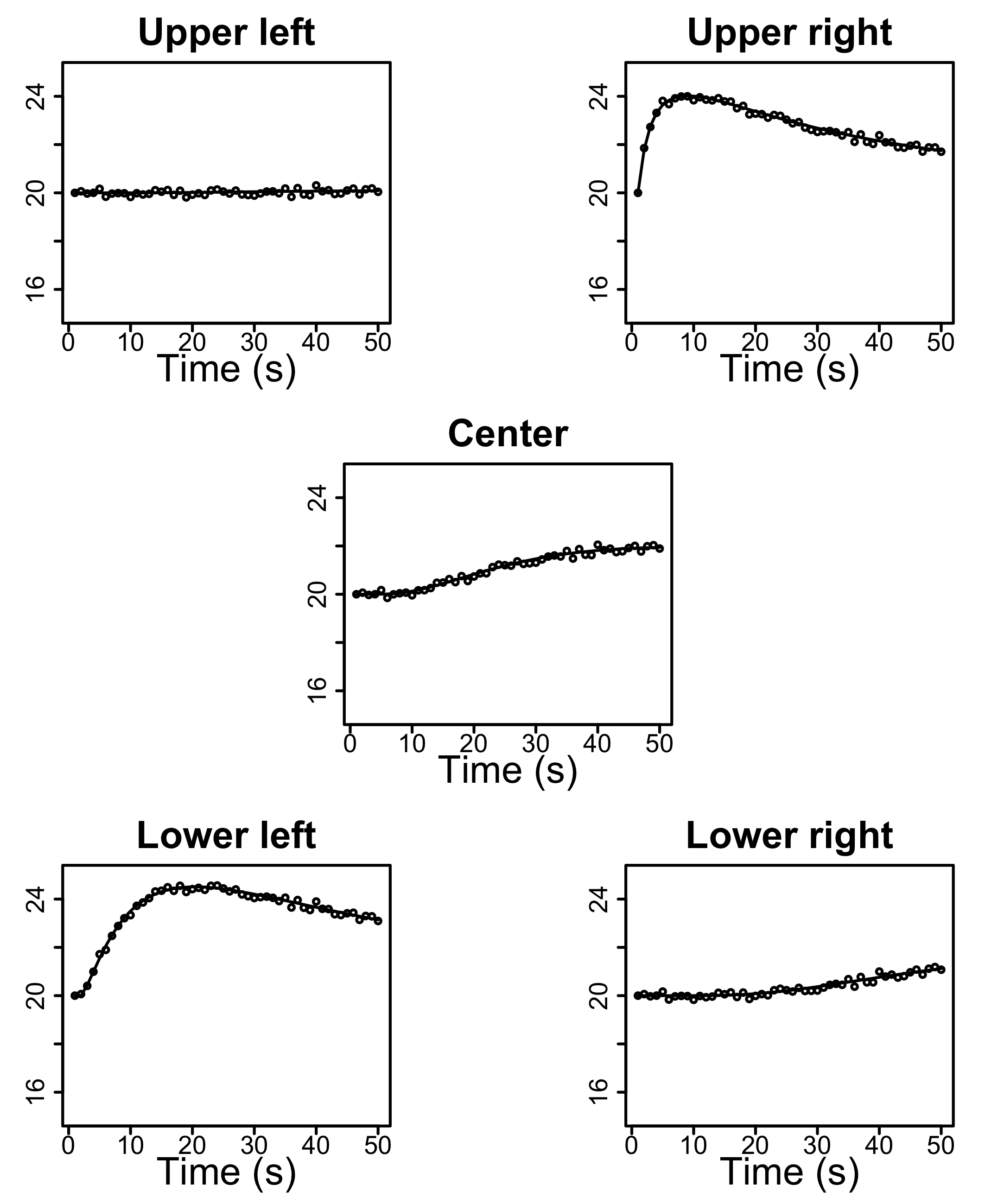}
	\caption{ Observations at the observation locations with the true curves for the two-event case}
	\label{supp4}
\end{figure}
\begin{figure}[!t]
	\centering
	\includegraphics[width=\textwidth]{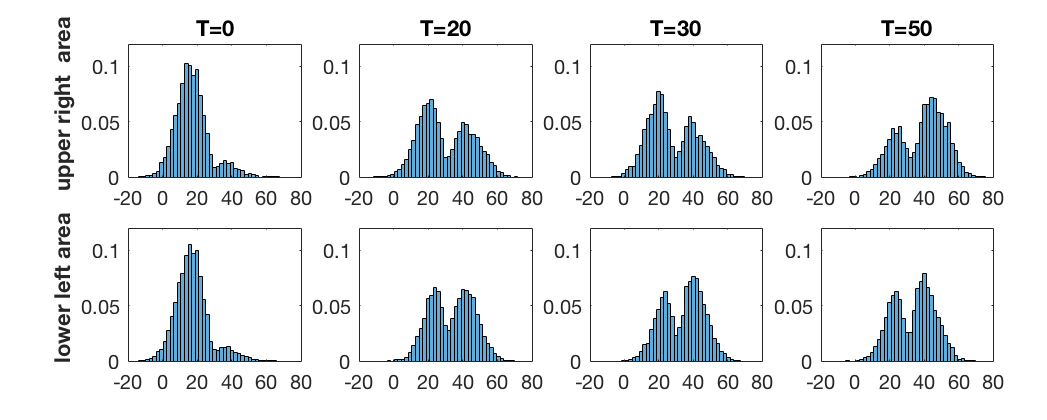}
	\caption{ Marginal pdfs at locations inside the events for increasing current time T for inversion with the selection Kalman model for the two-event case}
	\label{supp2}
\end{figure}
\begin{figure*}[!t]
	\centering
	\includegraphics[width=\textwidth]{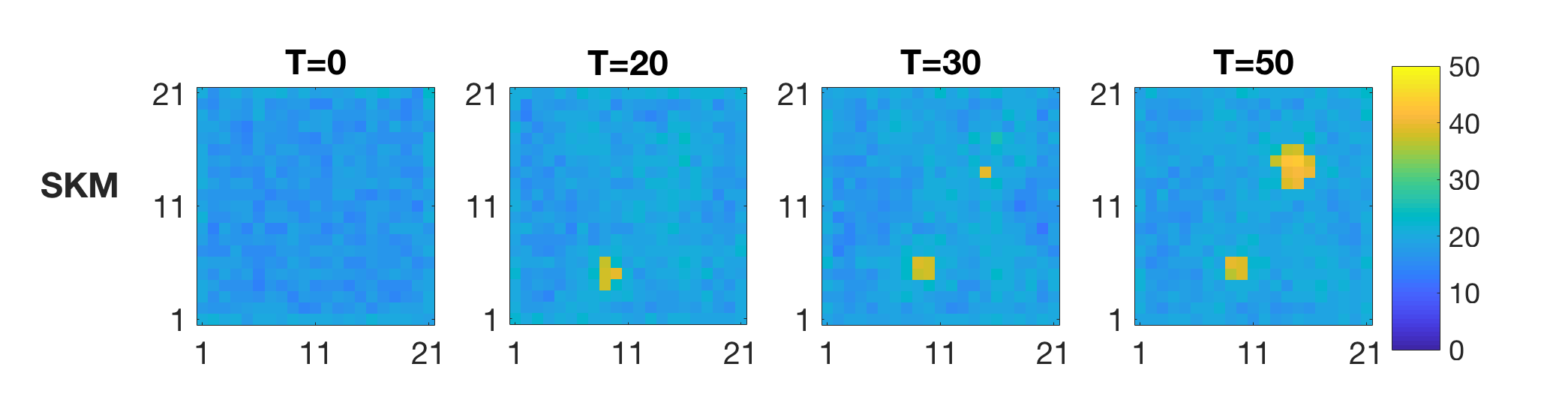}
	\caption{MMAP prediction at initial state for increasing current time T from the inversion with the selection Kalman model for the two-event case}
	\label{supp3}
\end{figure*}
\\
Figure \ref{finalreal1} and \ref{finalreal2} display realizations from the posterior pdf at $T=50$. For the selection Kalman model, see Figure \ref{finalreal1}, the event is precisely reproduced in the majority of realizations while for traditional Kalman model, see Figure \ref{finalreal2}, the event is only vaguely indicated. Note however that the realizations from the selection Kalman model reflect the bimodality of the prior model outside the central area where the five spot observation design provides  the most information. These observations are consistent with the results observed in Figure \ref{ressg} and \ref{resg}.\\
The computational demand for the selection Kalman model is considerably higher than for the traditional Kalman model, as the former requires sampling from high dimensional truncated Gaussian pdfs. The sampling becomes increasingly more resource consuming as the grid dimension increases. For $n=441$, as in our study, it only takes a few minutes to assess the posterior distribution $f(\vect{r}_{A}|\vect{d})$ on a regular laptop computer. When larger grid sizes are investigated, the MCMC algorithm may be parallelized  to reduce computational time.

In order to demonstrate the generality of the selection Kalman model, we defined an alternative true initial state with two extreme events, see Figure \ref{supp1}. We used exactly the same model parameters as in the primary case. Note in particular that the number of extreme events is not specified. The observed time series will of course be different, see Figure \ref{supp4}. These time series have many similarity with the ones from the primary case. We inspect the marginal pdfs at two monitoring locations, one inside each extreme event, as they evolve with current time T, see Figure \ref{supp2}. Both marginal pdfs are identical at current time T=0, and as current time T increases  the height of the high-value mode increases, indicating that both monitoring locations are within high-value events. In Figure \ref{supp3} the corresponding MMAP predictions are displayed for increasing current time T. We observe that location, areal extent and value of both extreme events are well reproduced, but not as well as for the single-event case since identifying two sources obviously is more complicated. The identification challenge is of course increasing with increasing number of extreme events.
\section{Conclusion}
We define a selection Kalman model based on a selection-Gaussian initial distribution and Gauss-linear dynamic and observation models. This model may represent spatial phenomena with initial states with spatial histograms that are skewed, peaked and multimodal. The selection Kalman model is demonstrated to be contained in the class of selection-Gaussian distributions and hence analytically tractable. The analytical tractability makes assessment of selection Kalman inversion fast and reliable. Moreover, an efficient recursive algorithm for assessing the selection Kalman model is specified. Note that the traditional Kalman model is a special case of the selection Kalman model, hence the latter can be seen as a generalization of the former. 
\\
A synthetic spatio-temporal case with an initial state including an extreme event  and Gauss-linear dynamic and observation models is used to demonstrate the characteristics of the methodology. We specify both a selection Kalman model and a traditional Kalman model, and evaluate their ability to restore the initial state based on the  observed time series. The time series are noisy observations  of the variable of interest collected at a set of sites. The selection Kalman model clearly outperforms the traditional Kalman model. The former model identifies location, areal extent and value of the extreme event very reliably. The traditional Kalman model only provides blurry indications with severe under-prediction of the extreme value. We conclude that for spatio-temporal variables where the initial spatial state have bimodal or multimodal spatial histograms, the selection Kalman model is far more suitable than the traditional Kalman model. 

The selection Kalman model has potential applications far beyond the simple case evaluated in this case study. For all spatio-temporal  problems  where multimodal spatial histograms  appear, the selection Kalman model should be considered. The model can easily be extended to a selection extended Kalman model, along the lines of the extended Kalman model. A more challenging and interesting extension  would be a selection ensemble Kalman model including  non linear dynamic and observation models. Research along these lines is currently taking place. 

\section*{Acknowledgment}
The research is a part of the Uncertainty in Reservoir Evaluation (URE) activity at the Norwegian University of Science and Technology (NTNU).

\section*{References}

\bibliography{ref}

\begin{appendices}
	\section{Recursive algorithm for assessing the traditional Kalman model}
	\label{mgf}
	\begin{alg}{1}{Joint Traditional Kalman Model}
		\label{alg:1}
		\begin{itemize} 
			\item Define
			\item[] $\vect{\mu}_t^r = \mathbb{E}[\vect{r}_{t}]$
			\item[] $\vect{\mu}_t^d =\mathbb{E}[\vect{d}_{t}]$
			%\vect{\Sigma}_t^u =& {\Cov}[\vect{r}_{0:t},\vect{d}_{0:t-1}]\\
			%\vect{\Sigma}_t^c =&{\Cov}[\vect{r}_{0:t},\vect{d}_{0:t}]\\
			\item[] $\mat{\Sigma}^{rr}_{ts} = \Cov(\vect{r}_{t},\vect{r}_s) = {\mat{\Sigma}^{rr}_{st}}^T$ 
			\item[] $\mat{\Sigma}^{dd}_{ts} = \Cov(\vect{d}_{t},\vect{d}_s) ={\mat{\Sigma}^{dd}_{st}}^T$ 
			\item[] $\mat{\Gamma}_{ts}^{rd}=\Cov(\vect{r}_{t},\vect{d}_s)={\mat{\Gamma}_{st}^{dr}}^T$
			\item Initiate
			\item[]$\vect{r}_0 \sim \varphi_n(\vect{r}_0;\vect{\mu}_0^r,\mat{\Sigma}_0^r)$
			\item[] $\vect{\mu}_0^r = \vect{\mu}_0^r$
			\item[]$\mat{\Sigma}_{00}^{rr}=\mat{\Sigma}_0^r$
			\item Iterate $t = 0,...,T$
			\begin{itemize}
				\item[]Likelihood model:
				%\item[]$f(\vect{r}_t,...\vect{r}_0,\vect{d}_0,...,\vect{d}_{t-1},\vect{d}_t)=N(\vect{\mu}_t^c,\mat{\Sigma}_t^c)$
				\begin{itemize}
					\item[]$\vect{\mu}_t^d =\mat{H}\vect{\mu}_t^r$
					%\item[]$\vect{\mu}_t^c = \begin{bmatrix}\vect{\mu}_t^u \\ \mat{H}(\vect{\mu}_t^u)_{1:n} \end{bmatrix}$
					\item [] $\mat{\Sigma}_{tt}^{dd}=\mat{H}\mat{\Sigma}^{rr}_{tt}\mat{H}^T+\mat{\Sigma}^{d|r}_t$
					\item[] Iterate $s= 0,...,t$
					\begin{itemize}
						\item[]$\mat{\Gamma}_{ts}^{rd}=\mat{\Sigma}^{rr}_{ts}\mat{H}^T$
					\end{itemize}
					\item[] End iterate s
				\end{itemize}
				\begin{itemize}
					\item[] If $t > 0$: Iterate $s= 0,...,t-1$
					\begin{itemize}
						\item[]$\mat{\Sigma}_{ts}^{dd}=\mat{H}\mat{\Gamma}_{ts}^{rd}$
					\end{itemize}
					\item[] End iterate s
				\end{itemize}
				%\item[]$\mat{\Sigma}_t^c = \begin{bmatrix} \mat{\Sigma}_t^u & %\mat{V}^T \\
				%    \mat{V} &  \mat{H}\mat{\Sigma}^{rr}_{tt}\mat{H}^T+\mat{\Sigma}^{d|r}_t\end{bmatrix}$
				\item[]Forwarding model:
				%\item[]$f(\vect{r}_{t+1},\vect{r}_t,...\vect{r}_0,\vect{d}_0,...,\vect{d}_t)=N(\vect{\mu}_{t+1}^u,\mat{\Sigma}_{t+1}^u)$
				\begin{itemize}
					\item[]$\vect{\mu}_{t+1}^r = \mat{A}_t\vect{\mu}_{t}^r$
					%\item[]$\vect{\mu}_{t+1}^u = \begin{bmatrix}   \mat{A}_t(\vect{\mu}_{t}^c)_{1:n} \\\vect{\mu}_{t}^c \end{bmatrix}$
					\item[] $\mat{\Sigma}_{(t+1)(t+1)}^{rr} = 	\mat{A}_t\mat{\Sigma}^{rr}_{tt}\mat{A}_t^T+\mat{\Sigma}^{r|r}_t$
					\item[] Iterate $s= 0,...,t$
					\begin{itemize}
						\item[]$\mat{\Sigma}_{(t+1)s}^{rr} = \mat{A}_t\mat{\Sigma}_{ts}^{rr}$
						\item[]$\mat{\Gamma}_{(t+1)s}^{rd}= \mat{A}_t\mat{\Gamma}^{rd}_{ts}$
					\end{itemize}
					\item[] End iterate s
				\end{itemize}
				
				%\item[]$\mat{\Sigma}_{t+1}^u = \begin{bmatrix}
				%	\mat{A}_t\mat{\Sigma}^{rr}_{tt}\mat{A}_t^T+\mat{\Sigma}^{r|r%}_t & \mat{W}^T \\
				%	\mat{W} &  \mat{\Sigma}_{c}^t
				%	\end{bmatrix}$
			\end{itemize}
			\item End iterate t
			\begin{align*}
			f \left( \begin{bmatrix}\vect{r} \\\vect{d} \end{bmatrix} \right)
			=& \varphi_{n(T+2)+m(T+1)}\left(\begin{bmatrix}\vect{r} \\\vect{d} \end{bmatrix}; \begin{bmatrix}
			\vect{\mu}_{{r}} \\
			\vect{\mu}_{d}
			\end{bmatrix},\begin{bmatrix}
			\mat{\Sigma}^{rr} & \mat{\Gamma}^{rd} \\
			\mat{\Gamma}^{dr} &  \mat{\Sigma}^{dd}
			\end{bmatrix}\right)
			\end{align*}
			is then  fully defined by the algorithm. 
		\end{itemize}
	\end{alg}
	\section{Parameters in the dynamic and likelihood models}
	\label{app:B}
	Dynamic $(n\times n)$-matrix A is derived from the following finite difference scheme: 
	\begin{align*}
	r_{i,j}^{t+1} =& r_{i,j}^{t} + \Delta t  (  -c_2 \frac{ r_{i,j+1}^{t+1} - r_{i,j}^{t+1} }{\Delta x} \\ +&\lambda \frac{r_{i+1,j}^{t+1}+ r_{i-1,j}^{t+1}+ r_{i,j+1}^{t+1}+ r_{i,j-1}^{t+1}+ r_{i,j}^{t+1} - 4 r_{i,j}^{t+1}}{\Delta x^2})
	\end{align*}
	Observation $(m\times n)$-matrix $\mat{H}$ is a binary selection matrix as:
	\begin{equation*}
	H_{i,j} = \begin{bmatrix} 
	0 &  \ldots &0 & 1 &  0 & \ldots  &  & & & & & & & \ldots & 0 \\
	0 &  \ldots & &\ldots & 0  & 1 & 0 & \ldots &  & & & &  & \ldots  & 0 \\
	0 &  \ldots & & & & \ldots& 0 & 1 & 0 & \ldots &  &  & & \ldots & 0 \\
	0 &  \ldots & & & & & &  \ldots& 0 & 1 & 0 & \ldots & & \ldots & 0 \\
	0 &  \ldots& &  & & & & &  & \ldots & 0& 1 & 0 & \ldots & 0 
	\end{bmatrix}
	\end{equation*}
\end{appendices}

\end{document}